\documentclass[longauth]{aa} 
\usepackage{graphicx}
\usepackage{amsmath}	% Advanced maths commands
\usepackage{amssymb}	% Extra maths symbols
\usepackage{multicol}        % Multi-column 
\usepackage{bm}		% Bold maths symbols, including upright Greek
\usepackage{pdflscape}	% Landscape pages
\usepackage{threeparttable} % To add links in your PDF file, use the package "hyperref" with options according to your LaTeX or PDFLaTeX drivers.
\usepackage[]{hyperref}
\usepackage{color}
\usepackage{txfonts}
\usepackage{longtable}
\usepackage{natbib,twoopt}
\usepackage{float}
\usepackage{subfig}

\bibpunct{(}{)}{;}{a}{}{,} %% natbib format like A&A and ApJ 

\newcommandtwoopt{\citeads}[3][][]{\href{http://adsabs.harvard.edu/abs/#3}%
{\citealp[#1][#2]{#3}}}
\newcommandtwoopt{\citepads}[3][][]{\href{http://adsabs.harvard.edu/abs/#3}%
{\citep[#1][#2]{#3}}}
\newcommandtwoopt{\citetads}[3][][]{\href{http://adsabs.harvard.edu/abs/#3}%
{\citet[#1][#2]{#3}}} 
\newcommandtwoopt{\citeyearads}[3][][]%
{\href{http://adsabs.harvard.edu/abs/#3}{\citeyear[#1][#2]{#3}}} 

\newcommand\kepler{\emph{{\it Kepler}}}
\newcommand\ktwo{\emph{{\it K2}}}
\newcommand\ms{$\mathrm{m\,s^{-1}}$}

\newcommand{\teff}{$T_{\rm eff}$}
\newcommand{\logg}{$\rm log\,g_\star$}

\newcommand{\Tzerob}[1][days]{$7738.82646 _{ - 0.00042 } ^ { + 0.00044 } $#1} 
\newcommand{\Pb}[1][days]{$1.208966 _{ - 0.000012 } ^ { + 0.000012 } $#1} 
\newcommand{\bb}[1][ ]{$0.21 _{ - 0.14 } ^ { + 0.23 } $#1} 
\newcommand{\dentrheeb}[1][${\rm g^{1/3}\,cm^{-1}}$]{$1.697 _{ - 0.128 } ^ { + 0.044 } $#1} 
\newcommand{\rrb}[1][ ]{$0.02323 _{ - 0.00037 } ^ { + 0.00058 } $#1} 
 
\newcommand{\mpb}[1][$M_{\oplus}$]{$3.74 _{ - 0.48 } ^ { + 0.50 } $#1} 
\newcommand{\rpb}[1][$R_{\oplus}$]{$1.62 _{ - 0.16 } ^ { + 0.17 } $#1} 
\newcommand{\ib}[1][deg]{$88.33 _{ - 2.10 } ^ { + 1.15 } $#1} 
\newcommand{\arb}[1][ ]{$7.23 _{ - 0.55 } ^ { + 0.19 } $#1} 
\newcommand{\ab}[1][AU]{$0.0210 _{ - 0.0026 } ^ { + 0.0024 } $#1} 
\newcommand{\insolationb}[1][${\rm F_{\oplus}}$]{$256 _{ - 23 } ^ { + 45 } $#1} 
\newcommand{\denstrb}[1][${\rm g\,cm^{-3}}$]{$4.89 _{ - 1.03 } ^ { + 0.39 } $#1} 
 
\newcommand{\denpb}[1][${\rm g\,cm^{-3}}$]{$4.81 _{ - 1.33 } ^ { + 1.97 } $\,#1} 
\newcommand{\grapb}[1][${\rm cm\,s^{-2}}$]{$1712 _{ - 354 } ^ { + 264 } $#1} 
\newcommand{\grapparsb}[1][${\rm cm\,s^{-2}}$]{$1395 _{ - 298 } ^ { + 391 } $#1} 
 
\newcommand{\Teqb}[1][K]{$1114 _{ - 26 } ^ { + 46 } $#1} 
\newcommand{\ttotb}[1][hours]{$1.281 _{ - 0.019 } ^ { + 0.020 } $#1} 

\newcommand{\Tzeroc}[1][days]{$7738.54961 _{ - 0.00145 } ^ { + 0.00146 } $#1} 
\newcommand{\Pc}[1][days]{$3.648227 _{ - 0.000119 } ^ { + 0.000117 } $#1} 
\newcommand{\bc}[1][ ]{$0.25 _{ - 0.16 } ^ { + 0.21 } $#1} 
 
\newcommand{\rrc}[1][ ]{$0.01820 _{ - 0.00041 } ^ { + 0.00054 } $#1} 
 
\newcommand{\mpc}[1][$M_{\oplus}$]{$1.47 _{ - 0.58 } ^ { + 0.59 } $#1} 
\newcommand{\rpc}[1][$R_{\oplus}$]{$1.27 _{ - 0.13 } ^ { + 0.13 } $#1} 
\newcommand{\ic}[1][deg]{$89.07 _{ - 0.92 } ^ { + 0.59 } $#1} 
\newcommand{\arc}[1][ ]{$15.10 _{ - 1.14 } ^ { + 0.39 } $#1} 
\newcommand{\ac}[1][AU]{$0.0439 _{ - 0.0055 } ^ { + 0.0050 } $#1} 
\newcommand{\insolationc}[1][${\rm F_{\oplus}}$]{$59 _{ - 5 } ^ { + 10 } $#1}

\newcommand{\denpc}[1][${\rm g\,cm^{-3}}$]{$3.87 _{ - 1.71 } ^ { + 2.38 } $\,#1} 
\newcommand{\grapc}[1][${\rm cm\,s^{-2}}$]{$1062 _{ - 461 } ^ { + 478 } $#1} 
\newcommand{\grapparsc}[1][${\rm cm\,s^{-2}}$]{$887 _{ - 369 } ^ { + 446 } $#1} 
 
\newcommand{\Teqc}[1][K]{$771 _{ - 18 } ^ { + 32 } $#1} 
\newcommand{\ttotc}[1][hours]{$1.825 _{ - 0.042 } ^ { + 0.042 } $#1} 

\newcommand{\Tzerod}[1][days]{$7740.96198 _{ - 0.00086 } ^ { + 0.00084 } $#1} 
\newcommand{\Pd}[1][days]{$6.201419 _{ - 0.000128 } ^ { + 0.000128 } $#1} 
\newcommand{\bd}[1][ ]{$0.864 _{ - 0.013 } ^ { + 0.022 } $#1} 
 
\newcommand{\rrd}[1][ ]{$0.02993 _{ - 0.00078 } ^ { + 0.00101 } $#1} 
 
\newcommand{\mpd}[1][$M_{\oplus}$]{$2.38 _{ - 0.69 } ^ { + 0.71 } $#1} 
\newcommand{\rpd}[1][$R_{\oplus}$]{$2.09 _{ - 0.21 } ^ { + 0.22 } $#1} 
\newcommand{\id}[1][deg]{$87.703 _{ - 0.253 } ^ { + 0.081 } $#1} 
\newcommand{\ard}[1][ ]{$21.50 _{ - 1.63 } ^ { + 0.56 } $#1} 
\newcommand{\ad}[1][AU]{$0.0625 _{ - 0.0078 } ^ { + 0.0071 } $#1} 
\newcommand{\insolationd}[1][${\rm F_{\oplus}}$]{$29 _{ - 3 } ^ { + 5 } $#1}

\newcommand{\denpd}[1][${\rm g\,cm^{-3}}$]{$1.42 _{ - 0.52 } ^ { + 0.75 } $\,#1} 
\newcommand{\grapd}[1][${\rm cm\,s^{-2}}$]{$641 _{ - 223 } ^ { + 225 } $#1} 
\newcommand{\grapparsd}[1][${\rm cm\,s^{-2}}$]{$534 _{ - 175 } ^ { + 219 } $#1} 
 
\newcommand{\Teqd}[1][K]{$646 _{ - 15 } ^ { + 26 } $#1} 
\newcommand{\ttotd}[1][hours]{$1.248 _{ - 0.033 } ^ { + 0.038 } $#1} 

\newcommand{\Tzeroe}[1][days]{$7739.87 _{ - 1.92 } ^ { + 1.96 } $#1} 
\newcommand{\Pe}[1][days]{$14.460 _{ - 0.106 } ^ { + 0.105 } $#1}

\newcommand{\qone}[1][]{$0.531 _{ - 0.089 } ^ { + 0.091 } $#1} 
\newcommand{\qtwo}[1][]{$0.398 _{ - 0.086 } ^ { + 0.087 } $#1} 
\newcommand{\uone}[1][]{$0.577 _{ - 0.125 } ^ { + 0.125 } $#1} 
\newcommand{\utwo}[1][]{$0.147 _{ - 0.126 } ^ { + 0.131 } $#1} 
\newcommand{\FIES}[1][${\rm m\,s^{-1}}$]{$31.77374 _{ - 0.00139 } ^ { + 0.00136 } $#1} 
\newcommand{\HARPS}[1][${\rm m\,s^{-1}}$]{$31.94794 _{ - 0.00037 } ^ { + 0.00036 } $#1} 
\newcommand{\HARPSN}[1][${\rm m\,s^{-1}}$]{$31.94888 _{ - 0.00034 } ^ { + 0.00035 } $#1} 
\newcommand{\jFIES}[1][${\rm m\,s^{-1}}$]{$1.25 _{ - 0.89 } ^ { + 1.55 } $#1} 
\newcommand{\jHARPS}[1][${\rm m\,s^{-1}}$]{$0.96 _{ - 0.39 } ^ { + 0.37 } $#1} 
\newcommand{\jHARPSN}[1][${\rm m\,s^{-1}}$]{$0.61 _{ - 0.40 } ^ { + 0.48 } $#1}

\def\kms{\hbox{\,km\,s$^{-1}$}}        %km.s-1
\def\vsini{\hbox{$v$\,sin\,$i_\star$}} %vsini
\begin{document}

\title{Mass determination of the 1:3:5 near-resonant planets transiting GJ\,9827 (K2-135)
\thanks{Based on observations made with a) the ESO-3.6m telescope at La Silla Observatory under programme ID 099.C-0491 and 0100.C-0808; b) the Italian Telescopio Nazionale Galileo operated on the island of La Palma by the Fundación Galileo Galilei of the Istituto Nazionale di Astrofisica; c) the Nordic Optical Telescope, operated by the Nordic Optical Telescope Scientific Association at the Observatorio del Roque de los Muchachos.}}

\author{J.~Prieto-Arranz\inst{1,2}%
   \and E.~Palle\inst{1,2}%
   \and D.~Gandolfi\inst{3}%
   \and O.~Barrag\'{a}n\inst{3}%
   \and E.W.~Guenther\inst{1,4}%
   \and F.~Dai\inst{5,6}%
   \and M.~Fridlund\inst{7,8}%
   \and T.~Hirano\inst{9}%
   \and J.~Livingston\inst{10}%
   \and P.~Niraula\inst{11}%
   \and C.\,M.~Persson \inst{8}%
   \and S.~Redfield\inst{11}%
%-------------Alphabetical-------------------
  \and S.~Albrecht\inst{12}%
  \and R.~Alonso\inst{1,2}%
  \and G.~Antoniciello\inst{3}%
  \and J.~Cabrera\inst{13}%
  \and W.\,D.~Cochran\inst{14}%
  \and Sz.~Csizmadia\inst{13}%
  \and H.~Deeg\inst{1,2}%
  \and Ph.~Eigm\"uller\inst{13}%
  \and M.~Endl\inst{14}%
  \and A.~Erikson\inst{13}%
  \and M.\,E.~Everett\inst{15}%
  \and A.~Fukui\inst{16}%
  \and S.~Grziwa\inst{17}%
  \and A.\,P.~Hatzes \inst{4}%
  \and D.~Hidalgo\inst{1,2}%
  \and M.~Hjorth\inst{12}%
  \and J.~Korth\inst{17}%
  \and D.~Lorenzo-Oliveira\inst{18}%
  \and F.~Murgas\inst{1,2}%
  \and N.~Narita\inst{10,19,20}%
  \and D.~Nespral \inst{1,2}%
  \and G.~Nowak\inst{1,2}%
  \and M.\,P\"atzold\inst{17}%
  \and P.~Monta\~nes Rodr\'iguez\inst{1,2}%
  \and H.~Rauer\inst{13,21}%
  \and I.~Ribas\inst{22,23}%
  \and A.\,M.\,S.~Smith\inst{13}%
  \and V.~Van~Eylen\inst{7}%
  \and J.\,N.~Winn\inst{5}%
}
\institute{
Instituto de Astrof\'\i sica de Canarias (IAC), 38205 La Laguna, Tenerife, Spain %1\\
\email{jparranz@iac.es}
\and 
Departamento de Astrof\'\i sica, Universidad de La Laguna (ULL), 38206, La Laguna, Tenerife, Spain %2
\and
Dipartimento di Fisica, Universit\'a di Torino, Via P. Giuria 1, I-10125, Torino, Italy %3
\and
Th\"uringer Landessternwarte Tautenburg, Sternwarte 5, 07778 Tautenburg, Germany %4
\and 
Department of Astrophysical Sciences, Princeton University, 4 Ivy Lane, Princeton, NJ 08544, USA %5
\and 
Department of Physics and Kavli Institute for Astrophysics and Space Research, Massachusetts Institute of Technology, Cambridge, MA 02139, USA %6
\and
Leiden Observatory, Leiden University, 2333CA Leiden, The Netherlands %7
\and
Department of Space, Earth and Environment, Chalmers University of Technology, Onsala Space Observatory, 439 92 Onsala, Sweden %8
\and 
Department of Earth and Planetary Sciences, Tokyo Institute of Technology, 2-12-1 Ookayama, Meguro-ku, Tokyo 152-8551, Japan %9
\and
Department of Astronomy, The University of Tokyo, 7-3-1 Hongo, Bunkyo-ku, Tokyo 113-0033, Japan %10
\and
Astronomy Department and Van Vleck Observatory, Wesleyan University, Middletown, CT 06459, USA %11
\and
Stellar Astrophysics Centre, Department of Physics and Astronomy, Aarhus University, Ny Munkegade 120, DK-8000 Aarhus C, Denmark %12
\and 
Institute of Planetary Research, German Aerospace Center, Rutherfordstrasse 2, 12489 Berlin, Germany %13
\and 
Department of Astronomy and McDonald Observatory, University of Texas at Austin, 2515 Speedway, Stop C1400, Austin, TX 78712, USA %14
\and
National Optical Astronomy Observatory, 950 North Cherry Avenue, Tucson, AZ 85719, USA %15
\and 
Okayama Astrophysical Observatory, National Astronomical Observatory of Japan, NINS, Asakuchi, Okayama 719-0232, Japan %16
\and 
Rheinisches Institut f\"ur Umweltforschung an der Universit\"at zu K\"oln, Aachener Strasse 209, 50931 K\"oln, Germany %17
\and
Universidade de S\~ao Paulo, Departamento de Astronomia do IAG/USP, Rua do Mat\~ao 1226, Cidade Universit\'aria, 05508-900 S\~ao Paulo, SP, Brazil %18
\and 
Astrobiology Center, NINS, 2-21-1 Osawa, Mitaka, Tokyo 181-8588, Japan %19
\and 
National Astronomical Observatory of Japan, NINS, 2-21-1 Osawa, Mitaka, Tokyo 181-8588, Japan %20
\and 
Center for Astronomy and Astrophysics, TU Berlin, Hardenbergstr. 36, 10623 Berlin, Germany %21
\and 
Institut de Ci\`encies de l\'~Espai (IEEC-CSIC), C/Can Magrans, s/n, Campus UAB, 08193 Bellaterra, Spain %22
\and
Institut d’Estudis Espacials de Catalunya (IEEC), E-08034 Barcelona, Spain %23
}

\date{Received dd mmm yyyy; accepted dd mmm yyyy}

\authorrunning{Prieto-Arranz et al.}  
\titlerunning{Mass determination of the 1:3:5 near-resonant planets transiting GJ\,9827 (K2-135)}

\abstract
% context heading (optional) 
{Multi-planet systems are excellent laboratories to test planet formation models, since all planets are formed under the same initial conditions. In this context, systems transiting bright stars can play a key role, since planetary masses, radii, and bulk densities can be accurately measured.}
% {heading} leave it empty if necessary
% aims heading (mandatory)
{GJ\,9827 (K2-135) has recently been found to host a tightly packed system consisting of three transiting small planets whose orbital periods of 1.2, 3.6, and 6.2 days are near the 1:3:5 ratio. GJ\,9827 hosts the nearest planetary system (d\,=\,$30.32\pm1.62$ pc) detected by \kepler\ and \ktwo . Its brightness (V\,=\,10.35\,mag) makes the star an ideal target for detailed studies of the properties of its planets.}
% Method (mandatory)
{Combining the \ktwo\ photometry with high-precision radial-velocity measurements gathered with the FIES, HARPS, and HARPS-N spectrographs we revise the system parameters and derive the masses of the three planets.}
% Results (mandatory)
{We find that GJ\,9827\,b has a mass of $M_\mathrm{b}=3.74^{+0.50}_{-0.48}$\,$M_\oplus$ and a radius of $R_\mathrm{b}=1.62^{+0.17}_{-0.16}$\,$R_\oplus$, yielding a mean density of $\rho_\mathrm{b}=\,$\denpb. GJ\,9827\,c has a mass of $M_\mathrm{c}=1.47^{+0.59}_{-0.58}$\,$M_\oplus$, radius of $R_\mathrm{c}=1.27^{+0.13}_{-0.13}$\,$R_\oplus$, and a mean density of $\rho_\mathrm{c}=\,$\denpc. For GJ\,9827\,d we derive $M_\mathrm{d}=2.38^{+0.71}_{-0.69}$\,$M_\oplus$\,, $R_\mathrm{d}=2.09^{+0.22}_{-0.21}$\,$R_\oplus$\,, and $\rho_\mathrm{d}=\,$\denpd.}
% Conclusions
{GJ\,9827 is one of the few known transiting planetary systems for which the masses of all planets have been determined with a precision better than 30\%. This system is particularly interesting because all three planets are close to the limit between super-Earths and mini-Neptunes. We also find that the planetary bulk compositions are compatible with a scenario where all three planets formed with similar core/atmosphere compositions, and we speculate that while GJ\,9827\,b and GJ\,9827\,c lost their atmospheric envelopes, GJ\,9827\,d maintained its atmosphere, owing to the much lower stellar irradiation. This makes GJ\,9827 one of the very few systems where the dynamical evolution and the atmospheric escape can be studied in detail for all planets, helping us to understand how compact systems form and evolve.}


\maketitle

\section{Introduction}
\label{sectI}

Systems containing multiple planets have drawn much attention because they have frequently been seen as potential Solar System analogues. However, none of the systems discovered so far resembles ours. The vast majority of multi-planet systems identified by the NASA’s \kepler\ space mission contains super-Earths ($1 \le R_\mathrm{p} \le 2 $\,$R_\oplus$) and mini-Neptunes ($2 \le R_\mathrm{p} \le 4$\,$\,R_\oplus$) in tightly packed configurations, with orbits smaller than the orbit of Mercury \citep{winn15}.  

Compact systems containing planets of different sizes and masses are the best test beds to constrain planetary formation mechanisms, since all planets have formed under the same initial conditions. The short orbital period increases the geometric probability to see the planets transiting their host stars, allowing us to measure the planetary radii. The Doppler reflex motion is larger, enabling the mass determination via radial velocity (RV) measurements using state-of-the-art, high-precision spectrographs. However, although more than 200 systems with three or more planets have been discovered so far, many questions remains unanswered.

How do compact planetary systems form? It has been proposed that planets with short orbital periods might have either formed \textit{in-situ} \citep{chiang13}, or at much larger distance from their host star and then moved inwards via type I or type II migration mechanisms \citep[for a review see][]{baruteau14}. Once the disk has been dispersed, planets could also migrate through planet-planet scattering \citep[see, e.g.,][]{marzari02}. Explaining the formation of compact systems with \textit{in-situ} formation is however not easy because a lot of material in the inner disk is required in order to form planets. Using an \textit{in-situ} formation model, \citet{hansen13} found that there are roughly 50\% more single-planet candidates observed than those produced by any model population. 

How can we observationally distinguish between different scenarios? In order to gain insights into the formation of compact systems, we have to understand whether the planets formed at large distance (e.g., beyond the snow-line), or close-in to their host star. It is now well accepted that the composition of a pre-main sequence disk -- where planet formation takes place -- depends on the radial distance from the host star. The chemical abundance of planets can thus be used to trace their formation. \citet{thiabaud15} showed that the C/O ratio is a good tracer to assess whether a given planet formed \textit{in-situ} or not. The Mg/Si and Fe/Si bulk composition ratios are also interesting tracers. In this respect, the discovery that the ultra-short period planet \object{K2-106 b} \citep{guenther17} has an iron core containing $80_{-30}^{+20}\%$ of its mass supports the fact that this planet might have formed in a metal rich environment -- typically close to the host star, where photophoresis process can separate iron from silicates in the early phase of planet formation \citep{wurm13}. On the contrary, if a close-in planet (a\,$\lessapprox$\,0.1\,AU) were found to have a high quantity of water, this would imply that the planet formed beyond the snow-line and then migrated inwards to its current position \citep{raymond08, lopez17}.

\begin{table}[!t]
  \centering 
  \caption{Equatorial coordinates, optical and near-infrared magnitude, and stellar parameters of GJ\,9827.}
  {\renewcommand{\arraystretch}{1.5}
  \begin{tabular}{lr}
    \hline
    \hline
    \multicolumn{2}{c}{GJ\,9827} \\
    \hline
    RA$^1$ (J2000.0)   &   23:27:04.83647 \\
	DEC$^1$ (J2000.0)  &  -01:17:10.5816 \\
	Distance$^1$ (pc)  &  $30.32\pm1.62$\\
	V-band magnitude$^2$ (mag)  &  $10.35 \pm 0.10$\\
	J-band magnitude$^3$ (mag)  &  $7.984 \pm 0.020$ \\
	Spectral type$^4$  &   K6\,V\\
	Effective temperature$^5$ \teff (K)  &  $4219\pm70$\\
    Surface gravity$^5$ \logg (cgs)  &  $4.650\pm0.050$\\
    Iron abundance$^5$ [Fe/H] (dex)  &  $-0.29\pm0.12$\\  
    Mass$^5$ $M_\star$ ($M_{\odot}$)  &  $0.650\pm0.060$\\
	Radius$^5$ $R_\star$ ($R_{\odot}$)  &  $0.637\pm0.063$\\
    Projected rot. velocity$^5$ \vsini\ (\kms)  &  $1.5\pm1.0$ \\
    Microturbulent velocity$^6$ $v_\mathrm{mic}$ (\kms) &  $0.9$ (fixed) \\
    Macroturbulent velocity$^7$ $v_\mathrm{mac}$ (\kms) &  $0.5$ (fixed)\\
    Interstellar reddening $A_\mathrm{v}$ (mag)$^5$ & $0.04\pm0.08$ \\

    \hline
  \end{tabular}}
\label{starpar}
\\
\flushleft
$^1$ Hipparcos, the New Reduction \citep{vanleeuwen07}. \\
$^2$ \citep{Mumford1956}. \\
$^3$ 2MASS \citep{2MASS}. \\
$^4$ \citet{houdebine16}. \\
$^5$ This work. \\
$^6$ \citet{bruntt10b} \\
$^7$ \citet{gray08}
\end{table}

On the other hand, as pointed out by \citet{izidoro17}, the period ratio distribution of planets in multi-planet systems can also provide some clues about the formation mechanisms involved. However, using N-body simulations together with a model of gaseous disc, \citet{izidoro17} found also that only 50-60\% of resonant chains became unstable whereas to match observations at least 75\% (and probably 90-95\% according to \kepler\ results) must be expected.

In order to address these questions, a well characterized sample of multi-planet systems transiting relatively bright stars, for which planetary radii, masses, and orbital parameters have been determined with high accuracy is needed. The three brightest systems known to host three or more planets for which masses have been determined for all planets, are \object{Kepler-89} (V=12.2 mag, 4 planets), \object{K2-32} (V=12.3 mag, 3 planets), and \object{Kepler-138} (V=12.9 mag, 3 planets). However, for most of the planets in these systems masses are known with a precision of only $\sim$50\% due to the faintness of the host stars.

To increase the sample of compact systems with planetary masses with a precision at least better than 30\%, we need to detect brighter systems (V$\lessapprox$12 mag) for which radial velocity (RV) precisions of 1\,\ms\ can be achieved using state-of-the-art spectrographs during a reasonable amount of telescope time.

\begin{table*}[!t]
\begin{center}
\caption{Spectroscopic parameters of GJ\,9827 as derived from the co-added HARPS (top) and HARPS-N (bottom) spectra using the two methods described in Sect~\ref{Sec:SpecAnalysis}.\label{Tab:spec_param}}
\begin{tabular}{lccccccc}
\hline
\hline
\noalign{\smallskip}
Method &  \teff\ & \logg\  &  [Fe/H] &  $R_\star$  & \vsini\ \\
       &    (K)  &  (cgs)  &   (dex) & ($R_\odot$) & (\kms) \\
\hline
\noalign{\smallskip}
\multicolumn{2}{l}{\emph{\bf HARPS}} \\
\noalign{\smallskip}
\texttt{SpecMatch-Emp} & 4203$\pm$70     & \dotfill      & $-$0.27$\pm$0.12       & 0.648$\pm$0.065 & \dotfill    \\
\texttt{SME 5.2.2}     & 4204$\pm$90     & 4.52$\pm$0.20 & $-$0.50$\pm$0.20       & \dotfill        & 1.5$\pm$1.0 \\
\hline
\noalign{\smallskip}
\multicolumn{2}{l}{\emph{\bf HARPS-N}} \\
\noalign{\smallskip}
\texttt{SpecMatch-Emp} & 4234$\pm$70     & \dotfill      & $-$0.30$\pm$0.12       & 0.651$\pm$0.065 & \dotfill    \\
\texttt{SME 5.2.2}     & 4236$\pm$90     & 4.44$\pm$0.20 & $-$0.53$\pm$0.20       & \dotfill        & 1.5$\pm$1.0 \\
\hline
\end{tabular}
\end{center}
\end{table*}

Using \ktwo\ time-series photometry from Campaign 12, we have recently discovered that the star GJ\,9827 -- also known as \object{K2-135} and \object{EPIC\,246389858} (Table~\ref{starpar}) -- hosts three transiting small planets ($R_\mathrm{p}\lesssim 2$\,$R_\oplus$) with orbital periods of 1.2, 3.6, and 6.2 days \citep{niraula17, Rodriguez2018}. With a distance of only $30.32\pm1.62$ pc, GJ\,9827 is the nearest planetary system detected by \kepler\ and \ktwo, and with V=10.35 mag (Table~\ref{starpar}) is the brightest system known to host 3 transiting planets.\\ 

In this paper, we present the high-precision RV measurements we collected between July and December 2017 to measure the masses of the three small planets transiting GJ\,9827. This work is part of the ongoing RV follow-up program of \ktwo\ transiting planets successfully carried out by our consortium \texttt{KESPRINT} \citep[see, e.g.,][]{nowak17,Fridlund2017,Gandolfi2017,barragan17,Dai2017,guenther17}.

\section{Ground based follow-up observations}
\label{sectII}

\subsection{High-spacial resolution}

\begin{figure}
\includegraphics[width=0.5\textwidth]{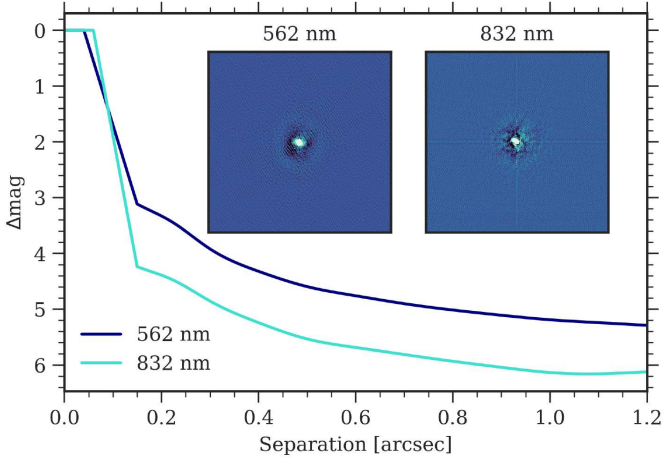}
\caption{Reconstructed images from WIYN/NESSI speckle interferometry and the resulting 5$\sigma$ contrast curves. The inset images are $4.6\arcsec\times4.6\arcsec$ and northeast is up and to the left.}
\label{speckle}
\end{figure}

We conducted speckle imaging observations of the host star with the WIYN 3.5-m telescope and the NASA Exoplanet Star and Speckle Imager (NESSI, \citet{scott16}, Scott et al., in prep.). The observations were conducted at 562nm and 832nm simultaneously, and the data were collected and reduced following the procedures described by \citet{howell11}. The resulting reconstructed images of the host star are $4.6\arcsec\times4.6\arcsec$, with a resolution close to the diffraction limit of the telescope. We did not detect any secondary sources in the reconstructed images, and we produced 5$\sigma$ detection limits from the reconstructed images using a series of concentric annuli (see Figure \ref{speckle}).

\subsection{FIES}

We collected 7 RV measurements of GJ\,9827 with the FIbre-fed Echelle Spectrograph \citep[FIES;][]{frandsen99, telting14} on the 2.56\,m Nordic Optical Telescope (NOT) at the Observatorio del Roque de los Muchachos, La Palma (Spain). The data have already been presented in \citet{niraula17}. We refer the reader to this work for a description of the observational strategy and data reduction. For the sake of completeness, we report the RV measurements in Table~\ref{tabRV1}.

\subsection{HARPS and HARPS-N}

We obtained 35 high-precision RVs with the HARPS spectrograph \citep{mayor03} on the 3.6\,m ESO telescope at La Silla Observatory under programs 099.C-0491 and 0100.C-0808, and 23 RV measurements with the HARPS-N spectrograph \citep{cosentino12} on the 3.58\,m Telescopio Nazionale Galileo (TNG) at La Palma under programs OPT17A\_64 and A36TAC\_12. The HARPS spectra were gathered from August 19 to October 24 2017 UT, and the HARPS-N spectra from July 29 to December 9 2017 UT. Both spectrographs have a resolving power of R$\,=\,\lambda/\Delta\lambda\,\approx\,115\,000$. HARPS covers the wavelength region from 3830\,\AA\ to 6900\,\AA, whereas HARPS-N from 3780\,\AA\ to 6910\,\AA. We used the second fiber of both instruments to monitor the sky background. All calibration frames were taken using the HARPS and HARPS-N standard procedures. The spectra were reduced and extracted using the dedicated data reduction software (DRS). The RVs were measured by cross-correlating the Echelle orders of the observed spectra with a K5 numerical mask \citep{baranne96,pepe02} and by fitting a Gaussian function to the average cross-correlation function (CCF). The DRS provides also the absolute RV, the bisector span (BIS) and full-width at half maximum (FWHM) of the CCF, and the Ca\,{\sc ii} S-index activity indicator. We list the HARPS and HARPS-N measurements in Tables~\ref{tabRV2} and~\ref{tabRV3}.

\section{Properties of the host star}
\label{sectIII}

\begin{figure*}
\includegraphics[width=1.0\textwidth]{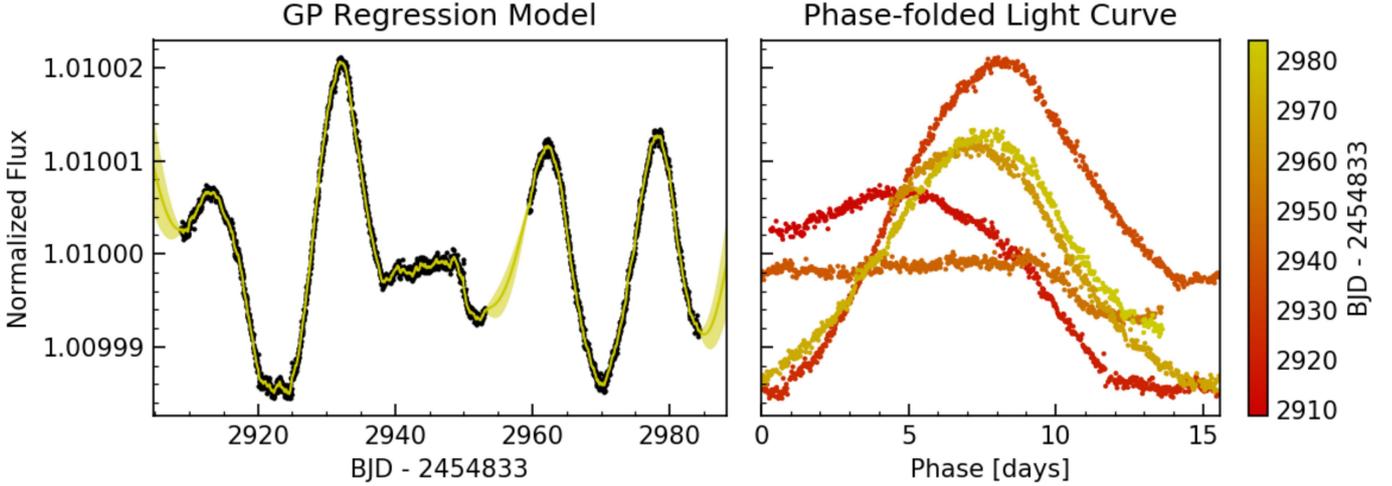}
\caption{\textit{Left panel}: Gaussian process regression model applied to the detrended K2 light curve. Black points are \ktwo\ light curve, yellow band is the Gaussian model. \textit{Right panel}: \ktwo\ detrended light curve phase-folded to the $P_{{\rm rot}/2}$ stellar rotational period.}
\label{gpmodel}
\end{figure*}

\subsection{Spectral analysis}
\label{Sec:SpecAnalysis}

In our previous paper \citep{niraula17}, we derived the spectroscopic parameters of GJ\,9827 using the co-added FIES spectrum, which has a S/N ratio of $\sim$150 per pixel at 5500\,\AA. As part of the analysis presented in this work, we refined the spectroscopic properties of the host star using the combined HARPS and HARPS-N spectra, taking advantage of their higher resolving power (R$\,\approx115\,000$) and S/N ratio ($\sim$440 and 400, respectively). The spectral analysis was performed following the same methods used in \citep{niraula17}, which, for the sake of completeness, are briefly described in the next paragraphs.

We used \texttt{SpecMatch-Emp} \citep{yee17}, a software suite that utilizes hundreds of Keck/HIRES template spectra of stars whose parameters have been accurately measured via interferometry, asteroseismology, spectral synthesis, and spectrophotometry. The fit is performed in the spectral region 5000-5900\,\AA. The output parameters of \texttt{SpecMatch-Emp}, namely, the effective temperature \teff, stellar radius $R_\star$, and iron abundance [Fe/H], are derived by interpolating those of the best matching library stars. Following \citet{hirano17}, prior to our analysis we reformatted the co-added HARPS and HARPS-N spectra so that they have the same spectral format as Keck/HIRES. 

\begin{figure*}[!th]
\centering
\includegraphics[width=\textwidth]{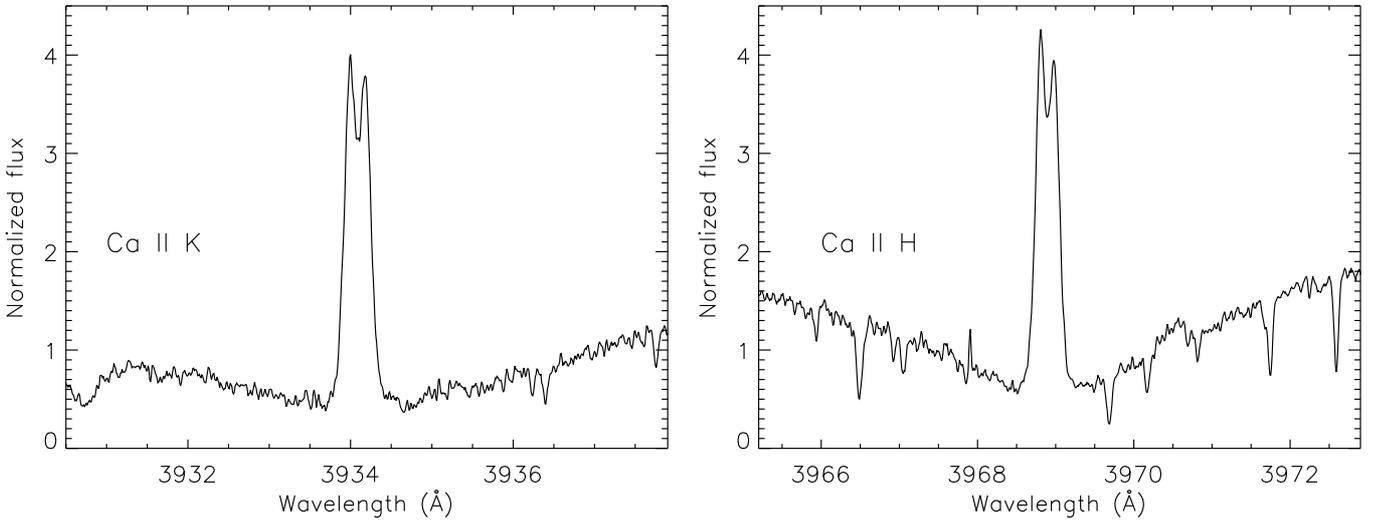}
\caption{Cores of the Ca\,{\sc ii} H\,\&\,K lines of GJ\,9827 as observed with HARPS.} 
\label{Fig:CaHK}
\end{figure*}

We also analyzed the HARPS and HARPS-N data with the spectral analysis package \texttt{SME} \citet{valenti96,valenti05}. \texttt{SME} calculates synthetic spectra from model atmospheres and fits them to the observed spectrum using a $\chi^2$ minimizing procedure. The analysis was carried out with the non-LTE version of the code (5.2.2) and \texttt{ATLAS\,12} model atmospheres \citep{kurucz13}. Following the calibration equation for Sun-like stars from \cite{bruntt10b}, we fixed the microturbulent velocity to $v_\mathrm{mic} =0.9$\kms. The macroturbulent velocity $v_\mathrm{mac}$ was assumed to be 0.5\kms \citep{gray08}. Following \citet{fuhrmann93, fuhrmann94}, the line wings of the H$_\mathrm{\alpha}$ and H$_\mathrm{\beta}$ lines were fitted to determine the effective temperature \teff. The surface gravity \logg\ was measured from the wings of the Ca\,{\sc i}~$\lambda$\,6102, 6122, 6162\,\AA\ triplet, and the Ca\,{\sc i} $\lambda$\,6439\,\AA\ line. The iron [Fe/H] and calcium [Ca/H] abundance, as well as the projected rotational velocity \vsini\ were derived fitting the profile of clean and unblended narrow lines in the spectral region between 6100 and 6500~\AA.  The analysis was finally checked with the Na doublet $\lambda$\,5889 and 5896\,\AA. 

We summarize our results in Table~\ref{Tab:spec_param}. The effective temperatures derived by \texttt{SpecMatch-Emp} and \texttt{SME} agree well within the nominal error bars. As for the iron abundance, the two methods provide consistent results within $\sim$2$\sigma$. It is worth noting that the error bars calculated by \texttt{SME} are larger than those given by \texttt{SpecMatch-Emp}, owing to the physical uncertainties of model atmospheres of cool stars (\teff\,$<$\,4500\,K). We therefore adopted the effective temperature and iron abundance measured by \texttt{SpecMatch-Emp} and averaged the estimates from the HARPS and HARPS-N spectra. For the projected rotational velocity \vsini, we adopted the value determined with \texttt{SME}. We found \teff\,=\,4219\,$\pm$\,70\,K, [Fe/H]\,=\,$-0.29$\,$\pm$\,0.12 (cgs), and \vsini\,=\,1.5$\pm$1.0\,\kms (Table~\ref{starpar}). The stellar radius and surface gravity were determined using a different method, as described in the following section.\\

\subsection{Stellar radius and mass}

We built the spectral energy distribution of GJ\,9827 using the Johnson B and V \citep{Mumford1956} and 2MASS JHKs \citep{2MASS} magnitudes. Following the method described in  \citet{Gandolfi2008}, we measured the interstellar redding ($A_\mathrm{v}$) along the line of sight to the star and found $A_\mathrm{v}=0.04\pm0.08$\,mag (Table~\ref{starpar}), which is consistent with zero, as expected given the proximity of GJ\,9827. We note that our result agrees with previous findings from \citet{McDonald2017} and \citet{Gontcharov2018}, confirming that the star suffers a negligible reddening.

We derived the stellar radius $R_\star$ by combining the Hipparcos'\,distance d\,=\,30.32\,$\pm$\,1.62\,pc \citep{vanleeuwen07}, with the apparent magnitude V\,=\,10.35\,$\pm$\,0.10 mag \citep{Mumford1956} and our effective temperature estimate \teff\,=\,4219\,$\pm$\,70\,K (Sect.~\ref{Sec:SpecAnalysis}). Assuming no reddening ($A_\mathrm{v}$=0 mag), we found a stellar radius of $R_\star$\,=\,$0.637$\,$\pm$\,0.063\,$R_\odot$, which agrees with the spectroscopic radius derived using \texttt{SpecMatch-Emp} (cfr. Table~\ref{Tab:spec_param}). 

We finally converted \teff, $R_\star$, and [Fe/H] into stellar mass $M_\star$ and surface gravity \logg\ using \citet{Mann2015}'s empirical equations coupled to Monte Carlo simulations. We found that GJ\,9827 has a mass of $M_\star$\,=\,0.650\,$\pm$\,0.060\,$M_\odot$ and a surface gravity of \logg\,=\,$4.650\,\pm\,0.050$~(cgs), which agrees with the spectroscopic gravity derived using \texttt{SME} (cfr. Table~\ref{Tab:spec_param}). According to our analysis performed with \texttt{SpecMatch-Emp}, the three stars\footnote{HIP\,12493, HIP\,97051, and HIP\,15095.} whose spectra best match the HARPS and HARPS-N spectra of GJ\,9827 have masses between 0.62 and 0.64\,$M_\odot$, confirming our results. The derived stellar mass and radius are are given in Table~\ref{starpar}.

\subsection{Stellar activity and rotation period}

The \ktwo\ light curve of GJ\,9827 displays a quasi-periodic photometric variability with a peak-to-peak amplitude of about 0.4\,\% (Fig.~\ref{gpmodel}, left panel). Given the late spectral type of the star (K6\,V), the observed photometric variation is very likely caused by active regions (sun-like spots and plages) crossing the visible stellar hemisphere as the star rotates about its axis. This is corroborated by the detection of emission components in the cores of the Ca\,{\sc ii} H \& K lines (Fig.~\ref{Fig:CaHK}), from which we measured an average S-index of $0.677\pm0.034$ and $0.739\pm0.021$ using the HARPS and HARPS-N spectra, respectively.

Applying the auto cross-correlation technique to the \ktwo\ light curve, \citep{niraula17} and \citep{Rodriguez2018} found that the rotation period $P_\mathrm{rot}$ of the star is either $\sim$17 or 30 days. We note that the ratio between the two measurements is close to 2, suggesting that the first might be the harmonic of the second. A visual inspection of the \ktwo\ light curve reveals that there are two dips whose minima occur at BJD$_\mathrm{TBD}\,-\,2454833$ \,$\approx$\,2922 and 2971 days, with a duration of $\sim$20 and 16 days, respectively (Fig.~\ref{gpmodel}, left panel). If the observed dips are caused by active regions crossing the visible hemisphere of GJ\,9827, the rotation period is likely longer than 17 days, suggesting that $P_\mathrm{rot}$ might be twice as long. A Gaussian process (GP) analysis of the \ktwo\ light curve (Sect.~\ref{Sect:GP}) shows a posterior bimodal distribution with rotational periods peaking at $15.1\pm1.6$ and $30.7\pm1.4$ days, showing just a minimal difference between both. Thus GP analysis does not provide a conclusive result about the rotation period of the star.

\begin{figure*}[!ht]
\centering
\includegraphics[width=0.98\textwidth]{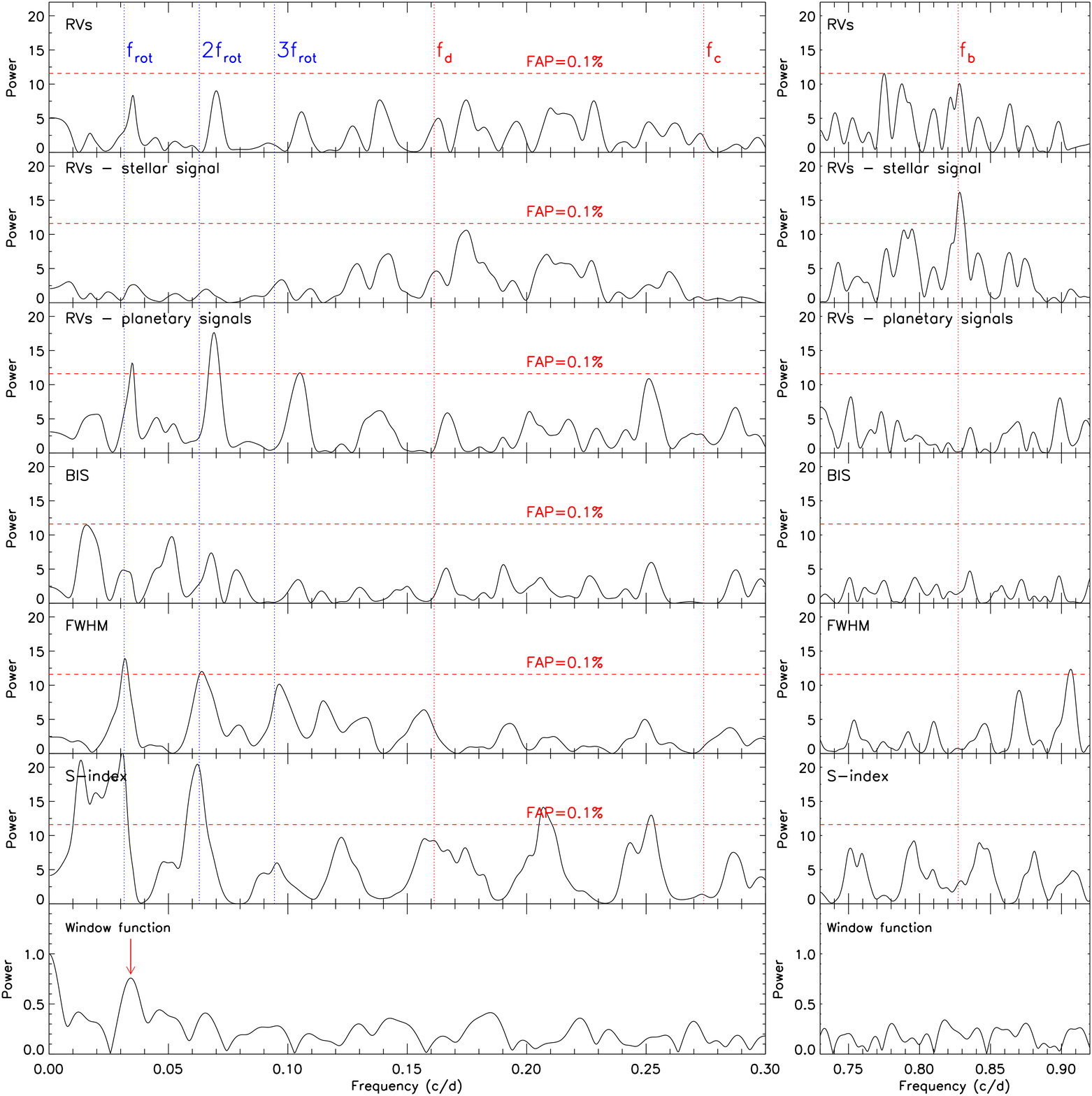}
\caption{Generalized Lomb-Scargle periodograms of the combined HARPS and HARPS-N datasets. The right and left columns cover two frequency ranges encompassing the 3 planetary signals (dotted vertical red lines), as well as the stellar rotation frequency and its first 2 harmonics (dotted vertical blue lines). From top to bottom: the RV data, the RV residuals after subtracting the signals of the 3 transiting planets, the RV residuals after subtracting the stellar activity signal, the BIS and FWHM of the CCF, and the window function. The dashed horizontal red lines mark the 0.1\,\% false alarm probabilities as derived using the bootstrap technique. The red arrow in the lower panel marks the peak discussed in the main text.}
\label{GJ9827_GLS}
\end{figure*}

\section{Frequency analysis of the HARPS and HARPS-N data}
\label{sec:freq}

The presence of active regions coupled to stellar rotation is expected to induce periodic and quasi-periodic RV signals at the stellar rotation frequency and its harmonics \citep[see, e.g.,][]{Hatzes2010,Haywood2015}. Using the code SOAP2 \citep{Dumusque2014}, we estimated the amplitude of the activity-induced RV signal -- the so-called activity-induced RV \emph{jitter} --  from the properties of the star, namely, its effective temperature, radius, rotation period, and photometric variability. We found that the predicted semi-amplitude of the RV jitter is $\sim$5\,\ms. Given the precision of most of our measurements ($\sim$1\,\ms), RV jitter is expected to be detected in our data-set.

We searched our Doppler time-series data for periodic signals associated with stellar activity by performing a frequency analysis of the RV measurements and activity indicators. For this purpose, we used only the HARPS and HARPS-N data because of the higher precision of the two data-sets. On epoch BJD=2458046, we purposely observed GJ\,9827 with both HARPS and HARPS-N nearly simultaneously (within less than 25 minutes) and used the two sets of measurements to estimate the RV, FWHM, BIS, and S-index offsets between the two instruments. We stress that these offsets have only been used to perform the periodogram analysis of the joint data.

Figure~\ref{GJ9827_GLS} displays the generalized Lomb-Scargle periodograms \citep[GLS;][]{Zechmeister2009} of the combined HARPS and HARPS-N data following the correction for instrument offset. From top to bottom, we show the periodograms of the combined HARPS and HARPS-N RVs, the RV residuals after subtracting the stellar activity signal assumed to be a Fourier component at $2f_\mathrm{rot}$ (Sect.~\ref{sectIV}), the RV residuals after subtracting the 3 planetary signals, the CCF bisector span (BIS), the CCF FWHM, the S-index, and the window function. Periodograms are displayed for two frequencies ranges encompassing the planetary and stellar signals. The vertical dotted lines mark the orbital frequencies of planet b, c, and d, as well as the stellar rotational frequency and its first 2 harmonics. The horizontal dotted lines mark the false alarm probabilities (FAP) of 0.1\% derived using the bootstrap method described in \citep{Kuerster1997}.

There are several important features to highlight in Figure~\ref{GJ9827_GLS}. The periodogram of the RV data shows peaks at the stellar rotational frequency and its harmonics (first row). The highest peak is found at about twice the rotation frequency with a semi-amplitude of $\sim$3\,\ms, in fairly good agreement with the value predicted by SOAP2 ($\sim$5\,\ms). Whereas the signals at the rotation frequency and its harmonics have a FAP\,$>$\,0.1 in the periodogram of the RV data (first panel), their significances increase with the FAP\,$\le$\,0.1 once the 3 planetary signals are subtracted from the time-series (third row). The periodograms of the CCF FWHM and S-index show also significant peaks (FAP\,$\le$\,0.1) whose frequencies are close to the stellar rotation frequency and its first harmonics, confirming that these signals are due to activity. 

The presence of two/three active regions separated by $\sim$180/120 degrees in longitude might account for the first and second harmonic of the fundamental rotation frequency. It's worth noting that the periodogram of the window function (lower row) shows a peak at 0.0342\,c/d ($\sim$29\,days; red arrow), reflecting the fact that our follow-up was carried out around new moon to avoid the contamination from the scattered Sun light. Since the sampling frequency is very close to the rotation frequency of the star, we acknowledge that the peaks associated to the rotation frequency and its harmonics might also arise from aliasing effects.

The periodogram of the RV residuals after subtracting the activity signal at $P_\mathrm{rot}/2$ (Sect.~\ref{sectIV}) shows a significant peak (FAP\,$\le$\,0.1) at the orbital frequency of GJ\,9827\,b (Figure~\ref{GJ9827_GLS}, second row). We conclude that the signal of the inner planet is clearly present in our RV data and that we would have been able to detect GJ\,9827\,b even in the absence of the K2 transit photometry.

\section{Data analysis}
\label{sectIV}

We modeled the \ktwo\ and RV data using two different techniques, as described in the following two sub-sections.

\subsection{Pyaneti analysis}
\label{Pyaneti_Analysis}

We performed the joint analysis to the photometric and RV data with the code \texttt{pyaneti} \citep{barragan17}, which explores the parameter space using a Markov chain Monte Carlo (MCMC) algorithm. We fitted Keplerian orbits to the RV data and used the limb-darkened quadratic transit model by \citet{mandel02} for the \ktwo\ transit light curves. In order to account for the \kepler\ long-cadence acquisition, we super-sampled the transit models using 10 subsamples per \emph{K2} exposure \citep{kipping10}. The fitted parameters and likelihood are similar to those used in previous analyses performed with \texttt{pyaneti} and described, e.g., in \citet{barragan16,Gandolfi2017}.

We fitted for a transit and a RV signal for each of the three planets. We sampled for $\rho_\star^{1/3}$ and recovered the scaled semi-major axis (a$_\mathrm{p}$/R$_\star$) of the three planets using Kepler's third law. We used uniform informative priors for all the parameters, except for the limb darkening coefficients for which we set Gaussian priors as described in \citet{niraula17}. 

As presented in the previous section, the RV data of GJ\,9827 shows activity-induced jitter at the stellar rotation frequency and its harmonics, with a semi-amplitude of $\sim$3\,\ms. The light curve of GJ\,9827 (Fig.~\ref{gpmodel}, left panel) suggests that the evolution time scale of active regions is longer than the \ktwo\ observations ($\sim$80 days). Since our FIES, HARPS, and HARPS-N RV follow-up covers $\sim$140 days, we can model the RV jitter using coherent sinusoidal signals at the stellar rotation frequency and its harmonics, similarly to the work described in, e.g., \citet{pepe13} and \citet{barragan17}.

In order to check which Fourier components at the rotation frequency and its harmonics can better describe the activity signal, we tested different RV models. The first model (3P) includes only the three planetary signals. The second model (3P+P$_\mathrm{rot}$) is obtained from 3P by adding a sinusoidal signal at the rotation period of the star ($P_{\rm rot}$$\sim$30 days). The third model called 3P+P$_\mathrm{\rm rot}/2$ includes three Keplerians and a sinusoidal signal at half the rotation period ($\sim$15 days). We also tested a model where two sinusoidal signals at both P$_{\rm rot}$ and P$_\mathrm{rot}/2$ were included. Since the stellar rotation period is not well constrained, we set uniform priors in the ranges $[P_{\rm rot}-2:P_{\rm rot}+2]$ and $[P_{{\rm rot}/2}-1:P_{{\rm rot}/2}+1]$.

Table~\ref{tab:models} summarizes out the results of our test, showing the goodness of the fit for each model. With the lowest Bayesian information criteria (BIC), the preferred model is 3P+P$_\mathrm{rot}/2$ (3 planets plus one sinusoidal signal at $\sim$15\,days). Table~\ref{tab:models} shows also that the semi-amplitudes of the three planetary signals do not change significantly when considering different models, providing evidence that the Doppler motion induced by the three planets is present in our RV data-set and does not depend on the Fourier components used to model the activity-induced RV signal. 

\begin{table*}%[!th]
\begin{center}
\caption{Model comparison.\label{tab:models}}
\begin{tabular}{ccccccrccr}
\hline
\hline
\noalign{\smallskip}
Model & $K_{\rm b}$ (m\,s$^{-1}$) & $K_{\rm c}$ (m\,s$^{-1}$) & $K_{\rm d}$ (m\,s$^{-1}$) & $K_{\rm rot}$ (m\,s$^{-1}$) & $K_{\rm rot/2}$ (m\,s$^{-1}$) & $\chi^2/\mathrm{dof}$  & BIC \\
\hline
\noalign{\smallskip}
    \noalign{\smallskip}
3P & $2.86 \pm 0.28$ & $0.80 \pm 0.24$ & $1.26 \pm 0.25$ & 0 & 0 & 2.8 & -500 \\
    \noalign{\smallskip}
3P + P$_{{\rm rot}}$ &  $2.96 \pm 0.30 $  & $1.11 \pm 0.27$ & $0.99 \pm 0.26$ & $5.68\pm0.84$ & 0 & 1.9 & -539 \\ 
    \noalign{\smallskip}
3P + P$_{{\rm rot}/2}$ &  $3.01 \pm 0.28$ & $0.85 \pm 0.27$ & $1.16 \pm 0.27$ & 0 & $3.18\pm0.38$ & 1.4 & -564 \\
    \noalign{\smallskip}
3P + P$_{{\rm rot}}$ + P$_{{\rm rot}/2}$  & $2.98\pm0.31$ & $0.82 \pm 0.27$ & $1.25\pm 0.30$ & $0.64^{+1.10}_{-0.47}$ & $3.27 \pm 0.50 $ & 1.7 & -488 \\
\hline
\end{tabular}
\end{center}
\end{table*}

\begin{figure*}
\centering
\subfloat{\includegraphics[width=0.50\textwidth]{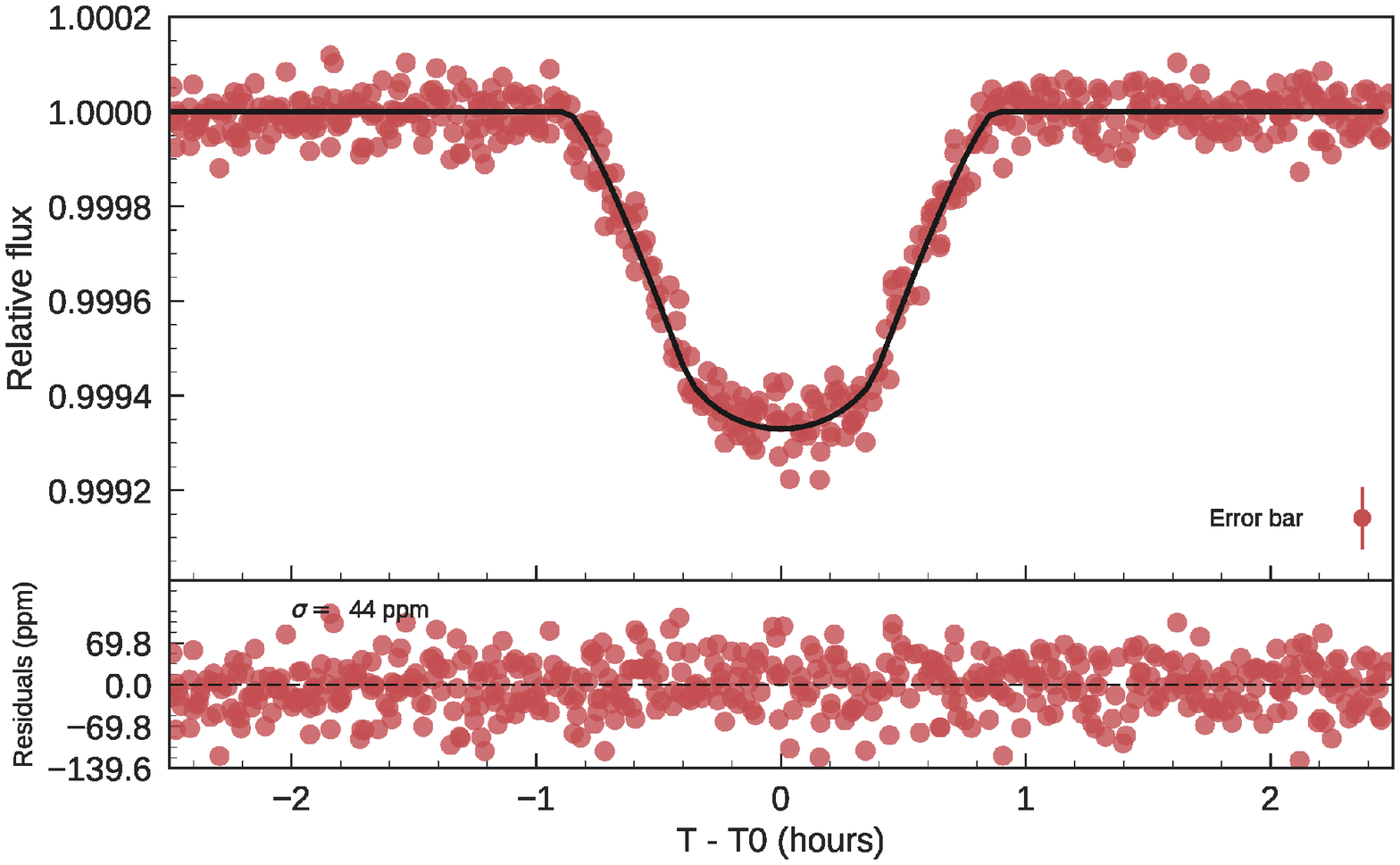}}
\subfloat{\includegraphics[width=0.50\textwidth]{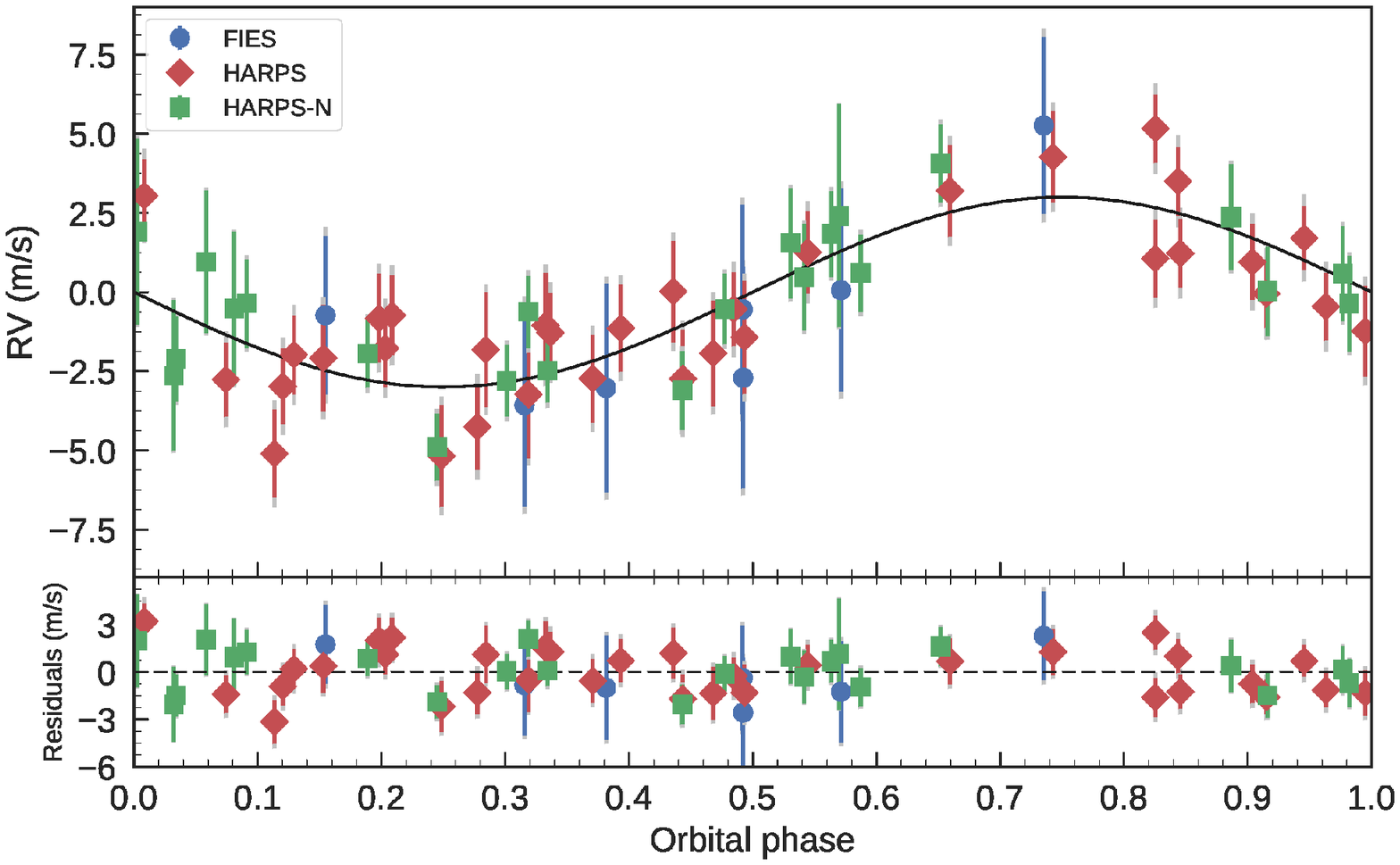}}\\
\subfloat{\includegraphics[width=0.50\textwidth]{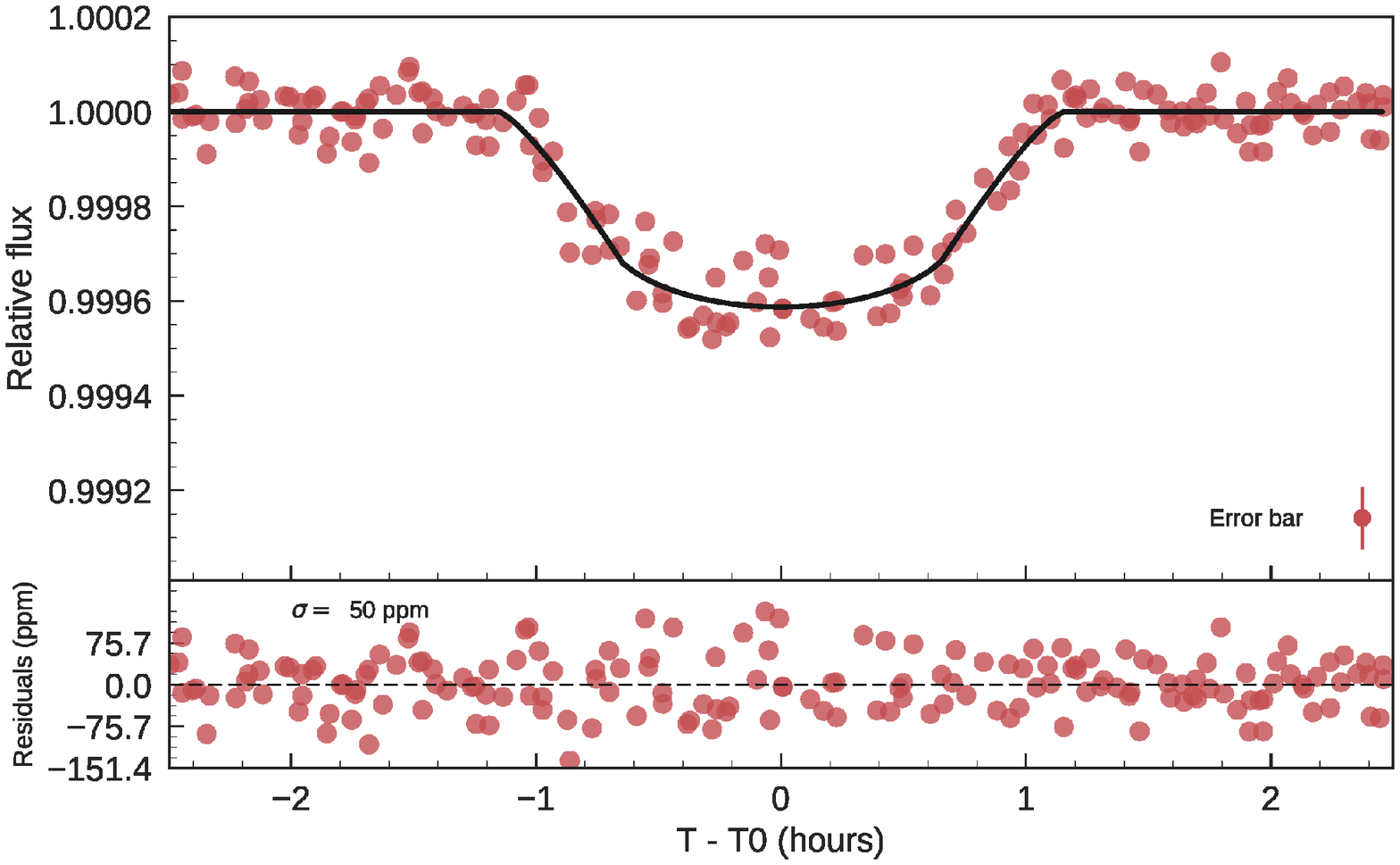}}
\subfloat{\includegraphics[width=0.50\textwidth]{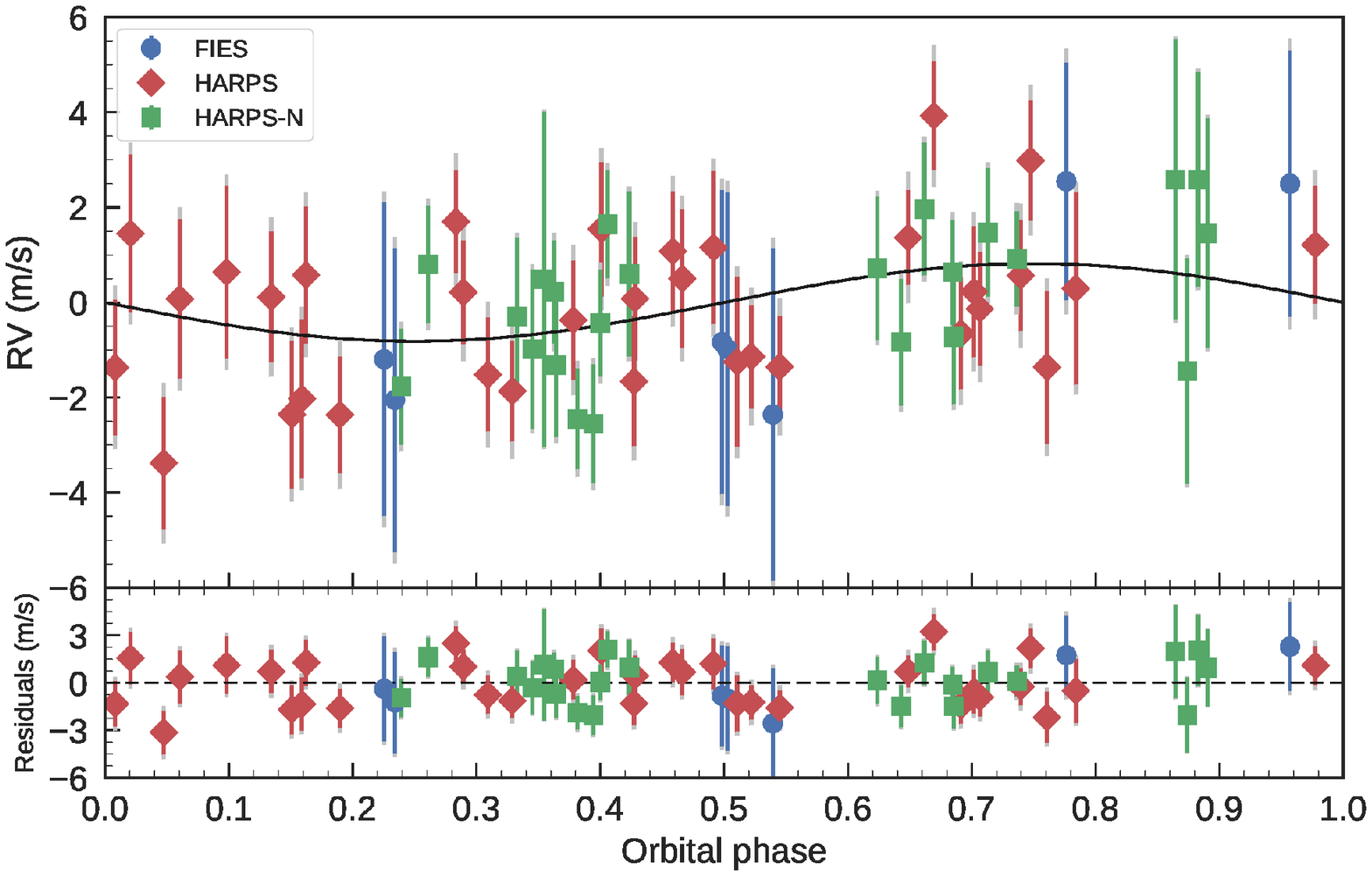}}\\
\subfloat{\includegraphics[width=0.50\textwidth]{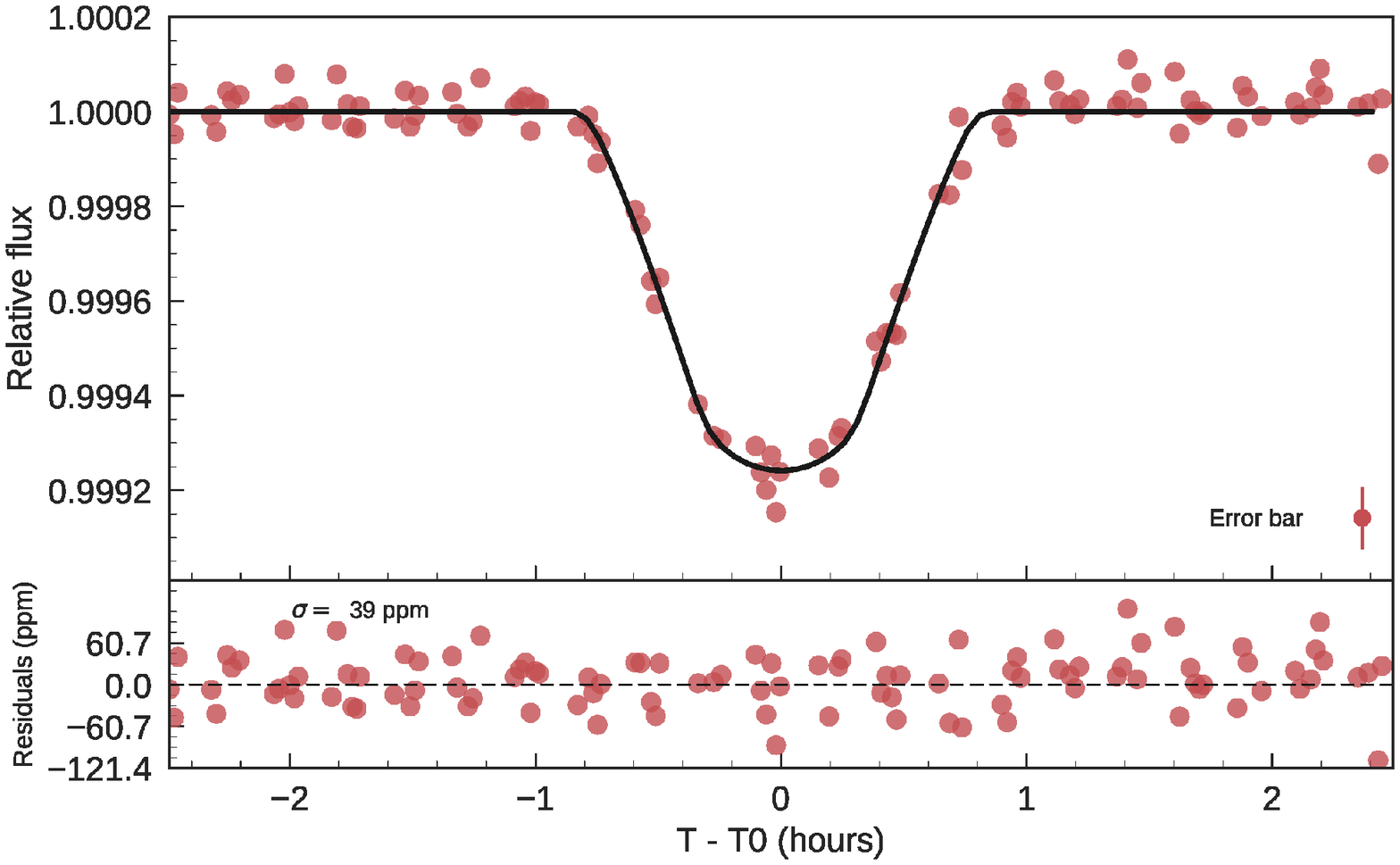}}
\subfloat{\includegraphics[width=0.50\textwidth]{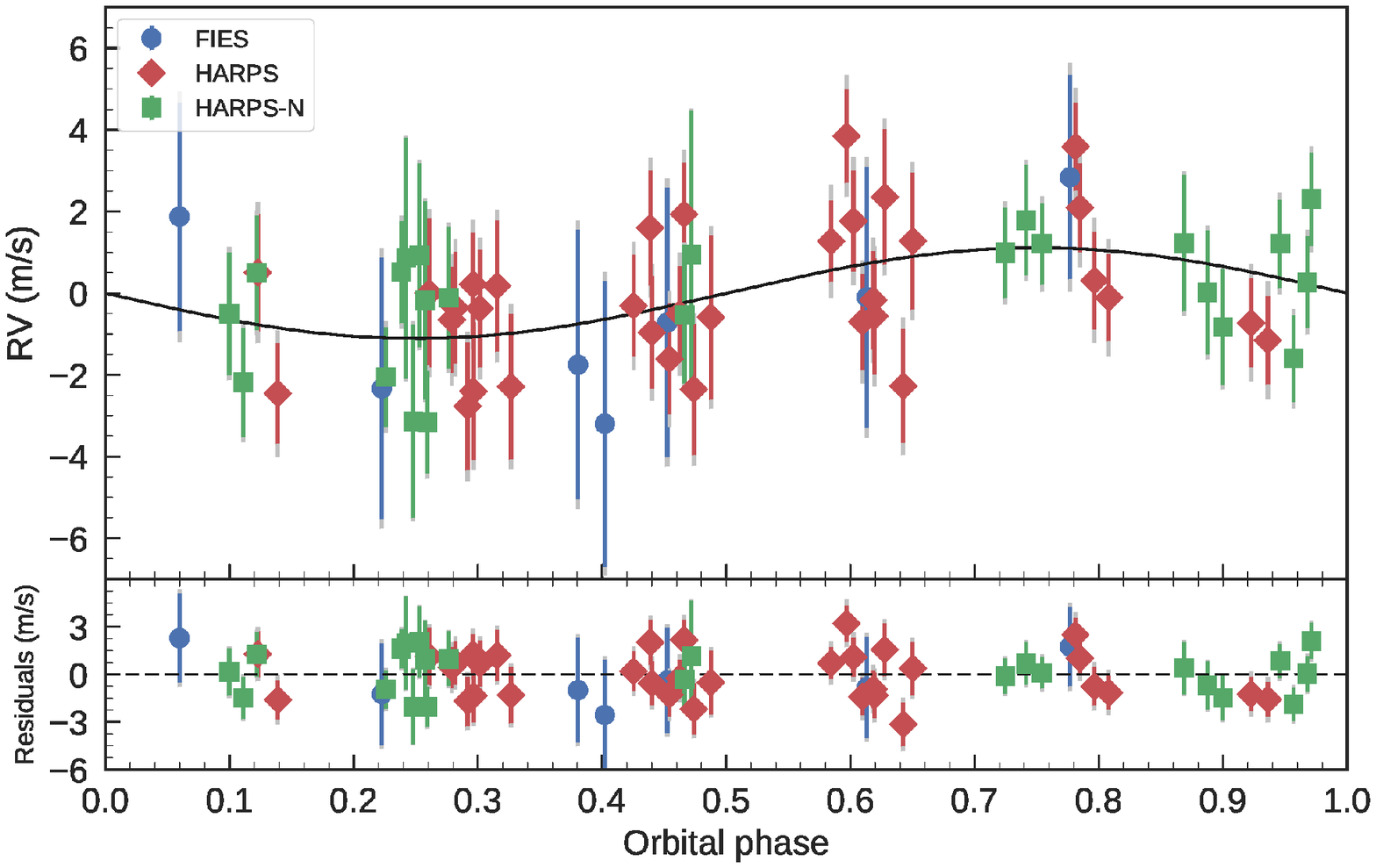}}\\
\caption{From top to bottom and left to right: transit fit and phase-folded RV curve of GJ\,9827\,b, GJ\,9827\,c, GJ\,9827\,d after removing the activity signal from the star and the signals from the other planets. The gray error bars account for additional instrumental noise and/or imperfect treatment of the various sources of RV variations.}
\label{GJ9827RV}
\end{figure*} 

\begin{figure}
\includegraphics[width=0.50\textwidth]{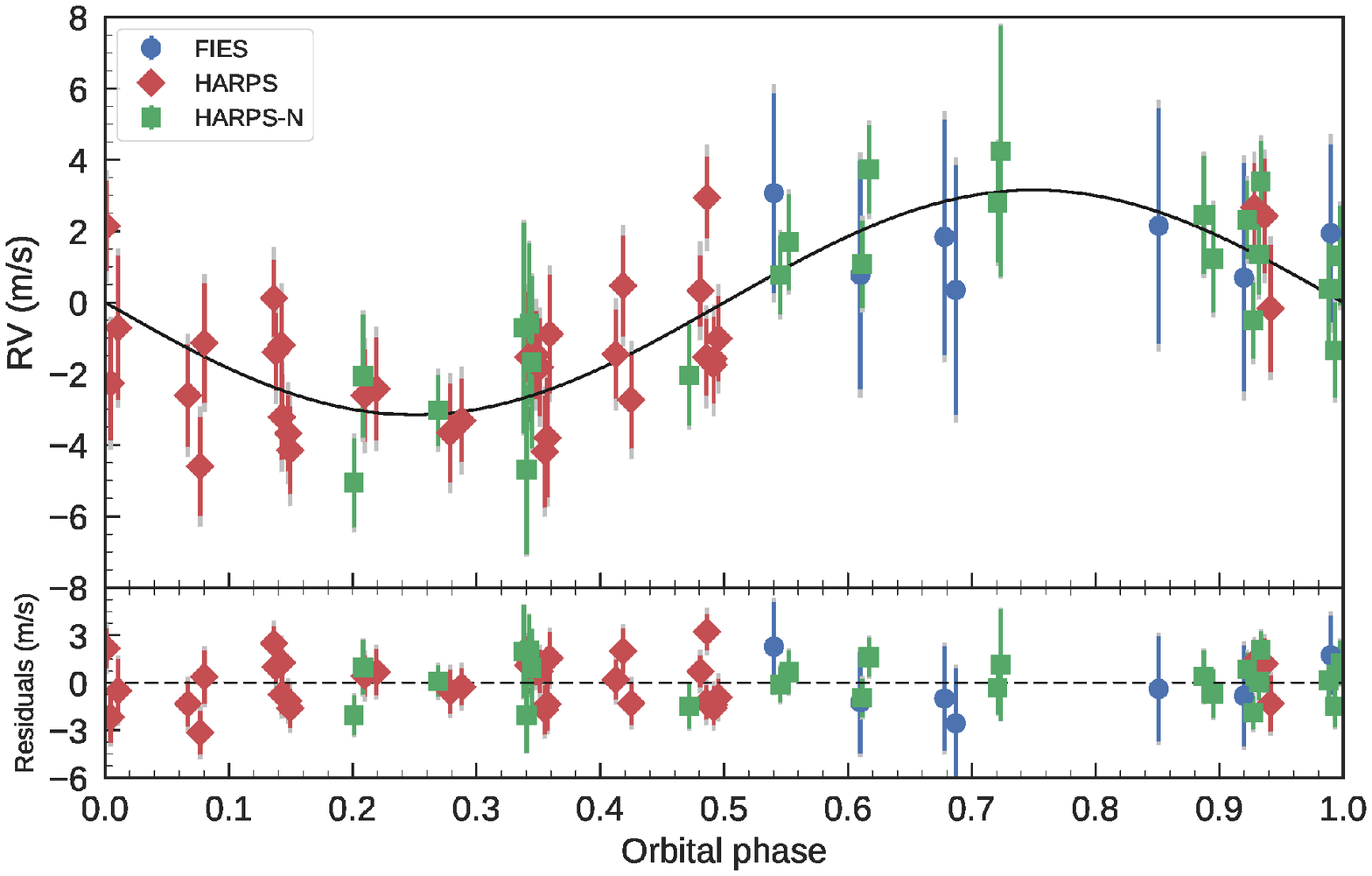}
\caption{RV curve of GJ\,9827 phase-folded to the first harmonic of the stellar rotation period ($P_\mathrm{rot}/2$\,=\,15.1\,days) after removing the signals of the three transiting planets.}
\label{RV_Stellar_Activity_Signal}
\end{figure} 

We performed a final joint analysis assuming that the RV data are described by the 3P+P$_{{\rm rot}/2}$ model. For the phase, amplitude, and period of the activity signal we adopted uniform priors. We included a jitter term for each spectrograph to account for additional instrumental noise not included in the nominal RV error bars and/or imperfect treatment of the various sources of RV variations. Since GJ\,9827 hosts a short-period multi-planetary system, we assumed tidal circularization of the orbits and fixed $e=0$ for all three planets \citep{vaneylen15}. We explored the parameter space with 500 Markov chains initialized at random positions in the parameter space. Once all chains converged, we ran 5000 iterations more. We used a thin factor of 10 to generate a posterior distribution of 250,000 independent points for each parameter. We used the posterior distribution of each parameter to infer their values and its uncertainty given the median and the 68.3\% credible interval. The final fits are shown in Fig.\ref{GJ9827RV} and Fig.~\ref{RV_Stellar_Activity_Signal}; parameter estimates are summarized in Table~\ref{parstable1}.

\subsection{Gaussian process}
\label{Sect:GP}

\begin{figure*}
\center
\includegraphics[width=1.0\textwidth]{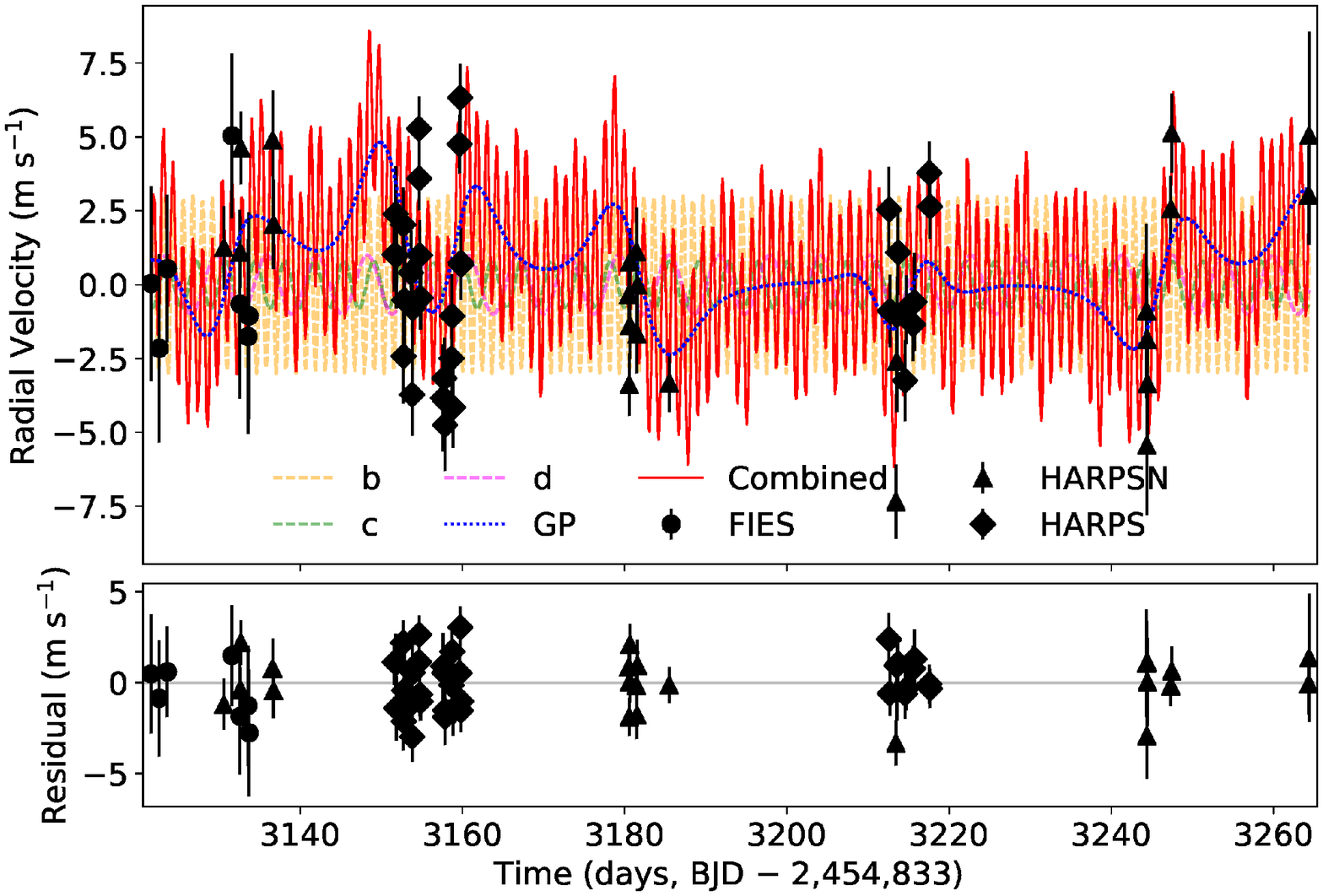} 
\caption{The measured radial velocity variation of GJ\,9827 from FIES (circles) and HARPS (diamonds) and HARPS-N (triangles). The red solid line is the best-fit model including the signal of the planets and the Gaussian Process model of the correlated stellar noise. The colored dashed line shows the signal of the planets. The blue dotted line shows the Gaussian Process model of correlated stellar noise.\label{fig:RV_GP} }
\end{figure*}

We also experimented with Gaussian Process (GP) to model the correlated RV noise associated with stellar activity. GP models stochastic processes with covariance matrices whose elements are generated by user-chosen kernel functions. GP regression has been successfully used to deal with the correlated stellar noise of the radial velocity datasets of several exoplanetary systems including \object{CoRoT-7}, \object{Kepler-78}, \object{Kepler-21}, and \object{K2-141} \citep{Haywood2014,Grunblatt2015,LM2016, Barragan2018}.

Our GP model was described in detail by \citet{Dai2017}. Briefly, we adopted a quasi-periodic kernel with the following hyperparameters: the covariance amplitude $h$, the correlation timescale $\tau$, the period of the covariance $T$, and $\Gamma$ which specifies the relative contribution between the squared exponential and periodic part of the kernel. For each of the transiting planets in GJ\,9827, we included a circular Keplerian signal specified by the RV semi-amplitude $K$, the orbital period $P_{\text{orb}}$ and the time of conjunction $t_{\text{c}}$. For each of spectrographs, we included a jitter parameter $\sigma$ and a systematic offset $\gamma$. We imposed Gaussian priors on $P_{\text{orb}}$ and $t_{\text{c}}$ with those derived from \ktwo\ transit modeling (Sect.\ref{Pyaneti_Analysis}). For the scale parameters $h$, $K$, and the jitter parameters we imposed Jeffreys priors. We imposed uniform priors on the systematic offsets $\gamma_{\text{HARPS}}$, $\gamma_{\text{HARPS-N}}$, and $\gamma_{\text{FIES}}$. Finally, for the hyperparameters $\tau$, $\Gamma$, and $T$ we imposed priors that were derived from a GP regression of the observed \ktwo\ light curve, as described below.

When coupled with stellar rotation, active regions on the host star give rise to quasi-periodic variations in both the measured RV and the flux variation. Given their similar physical origin, one would expect that GP with similar hyperparameters are able to describe the quasi-periodic variations seen in both datasets. Since the \ktwo\ light curve was measured with higher precision and sampling rate than our RV dataset, we trained our GP model on the \ktwo\ light curve. The resultant constrains on the hyperparameters were then used as priors when we analyzed the RV dataset. We adopted the covariance matrix and the likelihood function described by \citet{Dai2017}. We first located the maximum likelihood solution using the Nelder-Mead algorithm implemented in the {\tt Python} package {\tt scipy}. We then sampled the posterior distribution using the affine-invariant MCMC implemented in the code {\tt emcee} \citep{emcee}. We started 100 walkers near the maximum likelihood solution.  We stopped after running the walkers for 5000 links. We checked for convergence by calculating the Gelman-Rubin statistics which dropped below 1.03 indicating adequate convergence. We report the various parameters using the median and 16\%-84\% percentiles of the posterior distribution. The hyperparameters were constrained to be $\tau$\,=\,$6.1^{+4.0}_{-2.3}$ days, $T$\,=\,$15.1 \pm 1.6$ days and $\Gamma$\,=\,$0.77^{+0.47}_{-0.29}$. These served as priors in the subsequent GP analysis of the RV data. The GP model of the \ktwo\ light curve is shown in Fig.~\ref{gpmodel}.

In the analysis of the RV dataset with GP regression, we first found the maximum likelihood solution and sampled the parameter posterior distribution with MCMC using the same procedure as described above. The RV semi-amplitude for planet b, $K_b = 3.41\pm0.53$ m\,s$^{-1}$ was detected to a high significance. 
The RV signal of planet c was not securely detected in GP model. We therefore report the upper limit of $K_c\,<\,1.1$ m\,s$^{-1}$ at a 95\% confidence level. Finally, the RV signal of the outer planet was detected but with less confidence than the inner planet. We report a value of $K_d\,=\,1.06\pm0.52$ m\,s$^{-1}$. The amplitude of the correlated stellar noise is $h_{\text{rv}}\,=\,2.30^{+0.97}_{-0.66}$ m\,s$^{-1}$. All this values are in perfect agreement with the ones derived in previous section. Fig.~\ref{fig:RV_GP} shows the FIES, HARPS, and HARPS-N RVs of GJ\,9827 and the GP model. The planet parameter estimates are summarized in Table~\ref{parstable1}.
\linebreak

Given the good agreement between the results provided by the two methods and the fact that GP analysis provides only upper limit to the mass of the second planet, we adopted the values obtained with \texttt{Pyaneti}.

\section{Discussion}
\label{sectV}

\begin{figure}[ht]
\center
\includegraphics[width=0.5\textwidth]{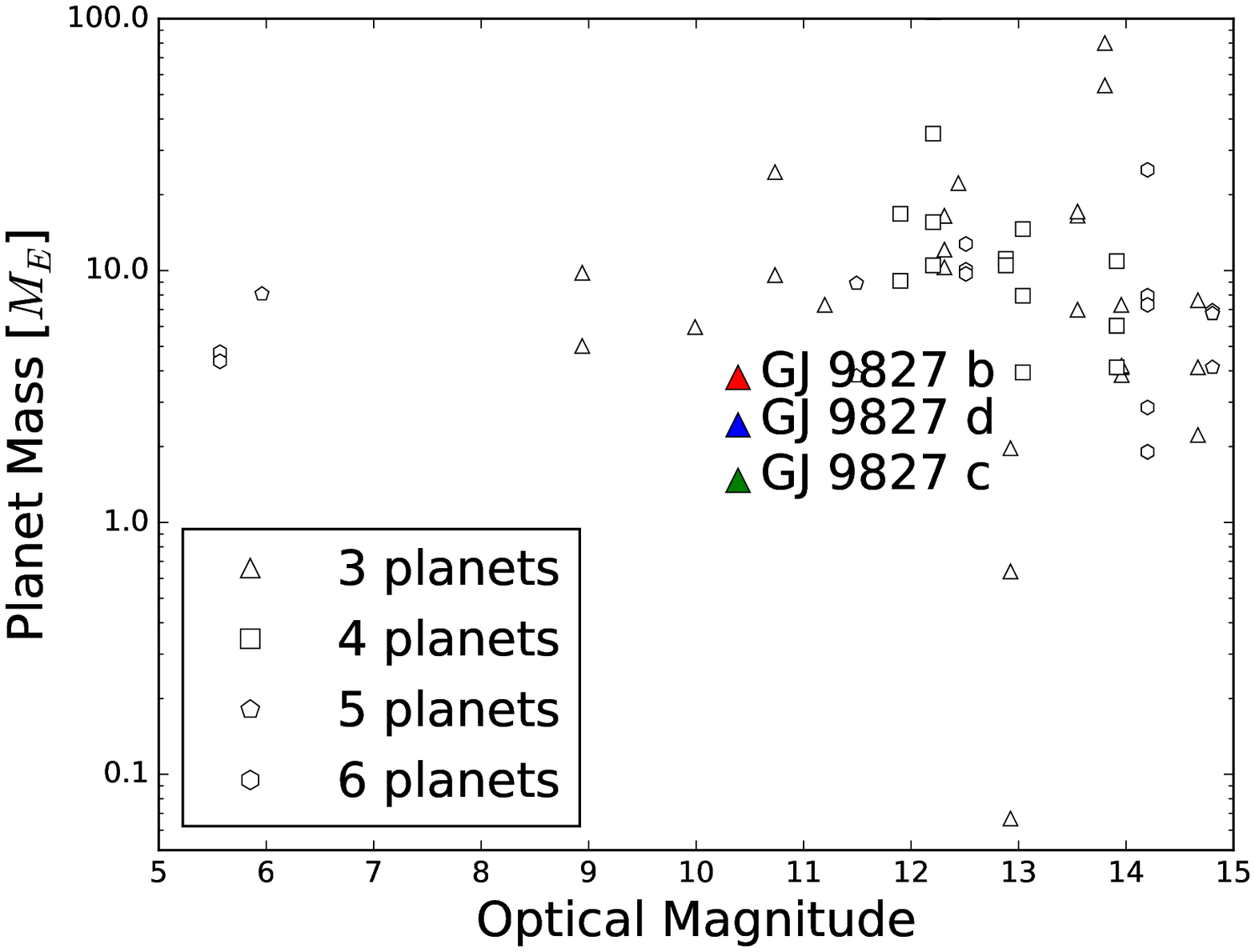}
\caption{Brightness-mass plot of planets with measured mass in multiple systems known to host at least three planets. With three transiting planets and V=10.35 mag, GJ\,9827 is the brightest multi-planet transiting system for which the masses of all planets have been measured. \label{AllKnownTripleSystems}}
\end{figure}

We determined masses, radii, and densities of the three planets known to transit GJ\,9827. We found that GJ\,9827\,b has a mass of $M_\mathrm{b}\,=\,3.74^{+0.50}_{-0.48}$\,$M_\oplus$ and a radius of $R_\mathrm{b}\,=\,1.62^{+0.17}_{-0.16}$\,$R_\oplus$, yielding a mean density of $\rho_\mathrm{b}=\,$\denpb. GJ\,9827\,c has a mass of $M_\mathrm{c}\,=\,1.47^{+0.59}_{-0.58}$\,$M_\oplus$, radius of $R_\mathrm{c}\,=\,1.27^{+0.13}_{-0.13}$\,$R_\oplus$, and a mean density of $\rho_\mathrm{c}\,=\,$\denpc. For GJ\,9827\,d we derived $M_\mathrm{d}=2.38^{+0.71}_{-0.69}$\,$M_\oplus$\,, $R_\mathrm{d}\,=\,2.09^{+0.22}_{-0.21}$\,$R_\oplus$\,, and $\rho_\mathrm{d}\,=\,$\denpd. Figure~\ref{AllKnownTripleSystems} shows the planetary masses as a function of the host star's visual magnitudes for systems known to host at least three planets. GJ\,9827 is the brightest (V=10.35 mag) transiting multi-planet system for which the masses of all planets have been measured.

In the next sub-sections we will address the following questions. What type of planets are GJ\,9827\,b, c, and d, and how well can we constrain their evolutionary history?

\subsection{Planets composition}
\label{sec:pltcomp}

To address these questions we can rely on the recent discovery of the existence of a bimodal distribution of planetary radii described by \citet{fulton17} and \citet{vaneylen17}. According to these works, there is a clear distinction between two different families of planets: super-Earths whose radius distribution peaks at $R_\mathrm{p}$\,$\sim$\,1.5\,$R_\oplus$, and sub-Neptunes whose radius distribution peaks at $R_\mathrm{p}$\,$\sim$\,2.5\,$R_\oplus$, separated by a dearth planet valley. The characteristics of this frontier (negative slope, dependence with period/incident flux) can be explained with photoevaporation of planetary atmospheres due to XUV radiation from the host stars.

\begin{figure}
\center
\includegraphics[width=0.5\textwidth]{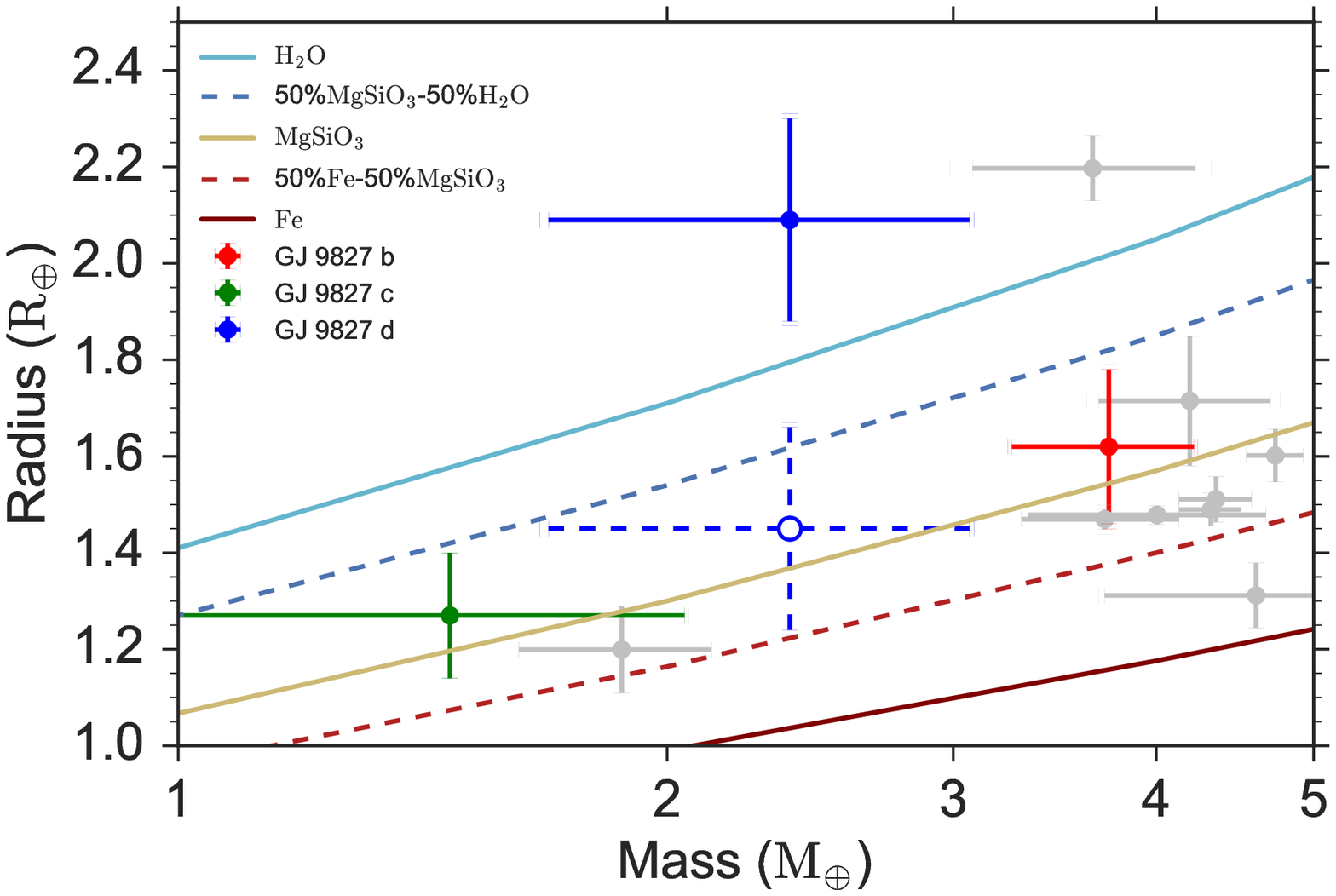} 
\caption{A mass-radius diagram for all rocky planets with masses between 1-5 $M_\oplus$ and radii between 1-2.5 $R_\oplus$, as registered in the TEPCat database. The solid circles indicate measurements of the mass and radius of the planets of GJ\,9827. The empty circle shows the inferred mass y radius of the nucleus of the third planet under the assumptions made on section \ref{sec:pltcomp}.
\label{fig:mr}}
\end{figure}

GJ\,9827 hosts a canonical terrestrial planet, GJ\,9827\,c, and two planets close to the dearth valley but from different sides: the super-Earth GJ\,9827\,b and the sub-Neptune GJ\,9827\,d. Fig.\,\ref{fig:mr} shows the position of the three planets in the mass-radius diagram along with the \citet{zeng16}'s theoretical models for different internal compositions. Planets b and c may have rocky nuclei with traces of lighter elements. Given its radius, planet d is likely surrounded by a large gaseous H/He-rich envelope. Since the innermost planets lie on the same isocomposition line of $\sim$80\%MgSiO$_3$-20\%H$_2$O (Fig.~\ref{fig:mr}), we can speculate that the outer planet might have a similar composition too. According to \citet{wolfgang15}, the atmosphere of GJ\,9827\,d would account for up to only $\sim$1\% of the total mass, yielding to a thickness of $\sim$0.6\,$R_\oplus$, i.e., $\sim$30\% of the planet's radius.

\subsection{Planets formation}

Based on the low abundance of resonant orbits among \kepler\ multi-planet systems, \citet{izidoro17} found that the instability rate of resonant chains is roughly 95\%. This means that GJ\,9827 belongs to the exclusive group of only $5\%$ of systems showing resonances. However, how this system came up to this configuration? To place GJ\,9827 in context, we show all transiting triple systems known so far in Table~\ref{tab:triple}, along with the ratios between the periods of their planets\footnote{Source: NASA Exoplanet Archive as of 1 February 2018.}. A plethora of these systems have 1:2 or 2:3 period ratios. These resonances have been theoretically predicted by \citet{wang17}, where type I migration plays a central role. Remarkably, the triple resonance 1:2:4 appears frequently where close-in terrestrial planets form driven by migration mechanisms \citep{sun17,wang17}. However, the resonant chain of the GJ\,9827 planetary system (1:3:5) is far more complex, indicating that possibly formation mechanism other than migration could be at play. 

How did GJ\,9827 reach the 1:3:5 resonance? According to \citet{izidoro17}, during planet formation, when the first embryo reaches the inner edge of the disk, its migration is stopped by the planet disk-edge interaction \citep{masset06} and other embryos migrate into a resonant chain. If this formation scenario is correct, several features would still be codified in the orbital eccentricity of the planets. As \citet{vaneylen15} demonstrated, from precise photometry (like the one gathered by \ktwo\ or by the upcoming space-telescope CHEOPS \citep{broeg13}) and using accurate asteroseismic density measurements (as those from the future PLATO mission \citep{PLATO}) the eccentricity of close-in planets could be precisely measured.

On the other hand, the masses of the three planets amount to a total mass of only $7.6\pm1.8\,M_\oplus$ (less than half the mass of Neptune), a quantity that could be compatible with an \textit{in-situ} formation scenario. \citet{chiang13} demonstrated that \textit{in-situ} formation in the minimum-mass extrasolar nebula is fast, efficient, and can reproduce many of the observed properties of close-in super-Earths. Therefore, if we could demonstrate that the three planets orbiting GJ\,9827 have formed \textit{in-situ} many information would be inferred about the primordial formation scenario of the system. One observationally testable property of close-in super-Earths mentioned by \citet{chiang13} is that they retain their primordial hydrogen envelopes. Additionally, if these planets did not migrate from the behind the snow-line and formed close to the host star they should not show any water features on their atmospheres. 

\begin{table}[ht]
  \centering 
  \caption{Triple transiting systems with measured masses}
  {\renewcommand{\arraystretch}{1.5}
  \begin{tabular}{lcrrr}
    \hline
    \hline
	System & Resonance & $M_1$ ($M_\oplus$) & $M_2$ ($M_\oplus$) & $M_3$ ($M_\oplus$) \\
    \hline
	Kepler-18    & 1:2:4 &  6.99 &  17.16 &  16.53 \\
    Kepler-30    & 1:2:4 & 11.44 & 638.83 &  23.20 \\
    Kepler-51    & 1:2:3 &  2.22 &   4.13 &   7.63 \\
    Kepler-60    & 3:4:5 &  4.19 &   3.85 &   4.16 \\
    Kepler-138   & 2:3:4 &  0.07 &   1.97 &   0.64 \\
    Kepler-289   & 1:2:4 &  7.31 &   4.13 & 133.49 \\
    K2-32        & 1:2:3 & 16.50 &  12.10 &  10.30 \\
    \bf{GJ\,9827}& 1:3:5 &  3.72 &   1.44 &   2.72 \\
    \hline
  \end{tabular}}
\label{tab:triple}
\\
$^1$ Data taken on 2018 Feb 1st, from NASA Exoplanet Archive: https://exoplanetarchive.ipac.caltech.edu
\\
\end{table}

\subsection{Planets atmosphere}

The fate of the atmosphere of an exoplanet strongly depends on the incident flux per surface unit due to photoevaporation processes. For GJ\,9827\,b, c, and d we calculated an incident flux relative to the Earth's of 256, 59 and 29, respectively. Interestingly although there is only a factor two between the flux of the second and third planet, the later seems to have a much lower density. This third planet lies well above the atmospheric loss frontier described in Figure~10 of \citet{vaneylen17}, while the other two are below. Moreover, the ratio between the incident fluxes and the masses of the planets are 70, 41 and 12, respectively. It is clear that the conditions of planet d are remarkably different from the other two. 

However, the low density of planet d seems to defy the photoevaporation models. With a mass of $3 M_\oplus$, previous models \citep{lammer03, owen16, wang17} would predict that planet d lost its H/He envelope within the first 100\,Myr of star's lifetime. We encourage more additional RV follow-up and transmission spectroscopy to pin down the properties of  planet d. The results may clarify our understanding of the photoevaporation process or unveil additional processes such as extreme out-gassing or late migration of planet. 

\begin{figure}[ht!]
\centering
\includegraphics[width=0.55\textwidth]{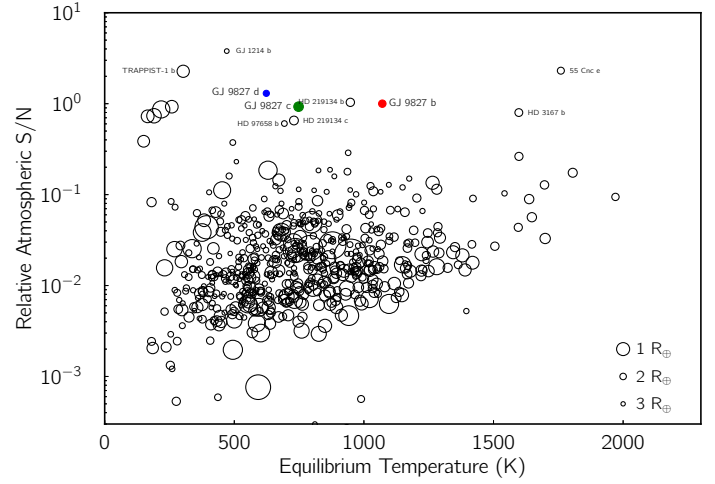} 
\caption{\label{fig:ATMsignal} The normalized atmospheric S/N for transiting planets with radii less than 3R$_\oplus$ as registered in the NASA Exoplanet Archive.}
\end{figure}

Given the brightness of the host star and small periods of the planets, the three planets transiting GJ\,9827 are excellent targets for atmospheric characterization using both space and ground-based facilities. \citet{niraula17} calculated the expected S/N of a planetary atmosphere using masses estimated by the mass-radius relationship by \citet{weiss14} and using a method similar to \citet{gillon2016}. Since we found that the masses are smaller than estimated from the mass-radius relation, these planets become even more attractive candidates for atmospheric studies than originally predicted. This is because the low surface gravity leads to a larger scale height, and thereby a larger atmospheric signal. GJ\,9827\,d ranks as the fourth best candidate overall (behind GJ\,1214\,b, 55\,Cnc\,e, and TRAPPIST-1\,b), and GJ\,9827\,b and c rank sixth and seventh, respectively, among the 601 transiting planets with radii $<$3R$_\oplus$, as shown in \autoref{fig:ATMsignal}. This makes the GJ\,9827 system a unique target for atmospheric studies.

\section{Conclusions}
\label{sectVI}

We have presented the characterization and mass determination of the three planets orbiting GJ\,9827 \citep{niraula17, Rodriguez2018}. \object{GJ\,9827} is a moderately active K6\,V star (S-index$\approx$0.7) with a rotational period of $P_\mathrm{rot}\approx30$ days transited by three small planets with masses of 3.74, 1.47, and 2.38\,$M_\oplus$, determined with a precision of $7.5\sigma$, $2.4\sigma$, and $3.4\sigma$, respectively. The system is an ideal laboratory to study planetary formation models and atmospheric photoevaporation. The densities of the three planets and the 1:3:5 orbital period ratio suggest an \textit{in-situ} formation scenario.

Our findings indicate that the third planet -- namely GJ\,9827\,d -- might have an extended atmosphere. The brightness of the host star (V=10.35 mag, J=7.984 mag) makes the transiting system around GJ\,9827 an ideal target to study the atmosphere of the three planets, using, for instance, JWST and ELT. By measuring the chemical abundances of the planetary atmospheres, it will be possible to further constrain the formation scenario of this system. Combining all this information, we will eventually unveil whether the planets formed roughly where they are found today, or whether they formed at much larger distance and then migrated inwards.

\begin{acknowledgements}

This work is partly financed by the Spanish Ministry of Economics and Competitiveness through projects ESP2014-57495-C2-1-R, ESP2016-80435-C2-2-R, and ESP2015-65712-C5-4-R of the Spanish Secretary of State for R\&D\&i (MINECO).
This project has received funding from the European Union's Horizon 2020 research and innovation programme under grant agreement No 730890. This material reflects only the authors views and the Commission is not liable for any use that may be made of the information contained therein.
DG gratefully acknowledges the financial support of the \emph{Programma Giovani Ricercatori -- Rita Levi Montalcini -- Rientro dei Cervelli (2012)} awarded by the Italian Ministry of Education, Universities and Research (MIUR).
SzCs, APH, MP, and HR acknowledge the support of the DFG priority program SPP 1992 "Exploring the Diversity of Extrasolar Planets" (HA 3279/12-1, PA 525/18-1, RA 714/14-1).
I.R. acknowledges support from the Spanish Ministry of Economy and Competitiveness (MINECO) and the Fondo Europeo de Desarrollo Regional (FEDER) through grant ESP2016-80435-C2-1-R, as well as the support of the Generalitat de Catalunya/CERCA programme.
This research has made use of the SIMBAD database, operated at CDS, Strasbourg, France.

\begin{table*}\footnotesize
\begin{center}
  \caption{Summary of the system parameters of GJ\,9827 determined in section \ref{sectIV} with both methods: Pyaneti and Gaussian Process. We adopt the former values for the Discussion section. \label{parstable1}}  
  \begin{tabular}{lcccc}
  \hline
  \hline
  \noalign{\smallskip}
  Parameter & GJ\,9827\,b & GJ\,9827\,c & GJ\,9827\,d & Sinusoidal signal \\
  \noalign{\smallskip}
  \hline
  \noalign{\smallskip}
  \noalign{\smallskip}
  \multicolumn{3}{l}{\emph{\bf Model Parameters: Pyaneti}} \\
  \noalign{\smallskip}
    Orbital period $P_{\mathrm{orb}}$ (days)  & \Pb[] & \Pc[] & \Pd[] & \Pe[] \\ 
    Transit epoch $T_0$ (BJD$_\mathrm{TDB}-$2\,450\,000)  & \Tzerob[] & \Tzeroc[] & \Tzerod[] & \Tzeroe[] \\  
    Scaled planet radius $R_\mathrm{p}/R_{\star}$ & \rrb[] & \rrc[] & \rrd[] & $\cdots$  \\
    Impact parameter, $b$ & \bb[] & \bc[] & \bd[] & $\cdots$   \\
	$\sqrt{e} \sin \omega_\star^{(a)}$  & 0 & 0 & 0 \\
    $\sqrt{e} \cos \omega_\star^{(a)}$  & 0 & 0 & 0 \\
        Doppler semi-amplitude variation $K$ (m s$^{-1}$) & $3.00\pm0.35$  &  $0.82\pm0.32$  &  $1.11\pm0.32$  &  $3.15\pm0.44$ \\

    Stellar density parametrization $\rho_{\star}^{1/3}$ (g$^{1/3}$ cm$^{-1}$)  & \dentrheeb[] \\
    Systemic velocity $\gamma_{\rm FIES}$ (km\,s$^{-1}$)  & \FIES[] \\
    Systemic velocity $\gamma_{\rm HARPS}$ (km\,s$^{-1}$)  & \HARPS[] \\
    Systemic velocity $\gamma_{\rm HARPS-N}$ (km\,s$^{-1}$)  & \HARPSN[] \\
   jitter $\sigma_{\rm FIES}$ (m\,s$^{-1}$)  & \jFIES[] \\
   jitter $\sigma_{\rm HARPS}$ (m\,s$^{-1}$)  & \jHARPS[] \\
   jitter $\sigma_{\rm HARPS-N}$ (m\,s$^{-1}$)  & \jHARPSN[] \\
    Parameterized limb-darkening coefficient $q_1^{(b)}$  &  \qone  \\
    Parameterized limb-darkening coefficient $q_2^{(b)}$  & \qtwo \\  
    \noalign{\smallskip}
    \hline
    \noalign{\smallskip}
    \multicolumn{3}{l}{\emph{\bf Derived Parameters: Pyaneti}} \\
    \noalign{\smallskip}
    Planet mass $M_\mathrm{p}$ ($M_{\rm \oplus}$) &  \mpb[] &  \mpc[] &  \mpd[] &  $\cdots$  \\
    Planet radius $R_\mathrm{p}$ ($R_{\rm \oplus}$)  & \rpb[]  & \rpc[] & \rpd[] & $\cdots$ \\
    Planet density $\rho_{\rm p}$ (g\,cm$^{-3}$)  & \denpb[] & \denpc[] &\denpd[] & $\cdots$ \\
    Surface gravity $g_{\rm p}$ (cm\,s$^{-2}$)  & \grapparsb[] & \grapparsc[] & \grapparsd[] \\
    Surface gravity$^{(c)}$ $g_{\rm p}$ (cm\,s$^{-2}$)  & \grapb[] & \grapc[] & \grapd[] \\
	Scaled semi-major axis $a/R_\star$  & \arb[] & \arc[] & \ard[] & $\cdots$ \\
	Semi-major axis $a$ (AU)  & \ab[] & \ac[]& \ad[]& $\cdots$  \\

    Orbit inclination $i_\mathrm{p}$ ($^{\circ}$) & \ib[] & \ic[] & \id[] & $\cdots$ \\
    Transit duration $\tau_{14}$ (hours) & \ttotb[] & \ttotc[] & \ttotd[] \\
    Equilibrium temperature$^{(\mathrm{d})}$  $T_\mathrm{eq}$ (K)  & \Teqb[] & \Teqc[] & \Teqd[] \\
        Insolation  $F$ ($F_{\oplus}$)  & \insolationb[] & \insolationc[] & \insolationd[] \\
    Stellar density (from light curve) & \denstrb[] \\
    Linear limb-darkening coefficient $u_1$ & \uone \\
    Quadratic limb-darkening coefficient $u_2$ &  \utwo\\
    \noalign{\smallskip}
\hline
\noalign{\smallskip}
\multicolumn{3}{l}{\emph{\bf Model Parameters: Gaussian Process}} \\
\noalign{\smallskip}
Doppler semi-amplitude variation $K$ (m s$^{-1}$) & $3.41\pm0.53$  &  $<1.10$  &  $1.06\pm0.52$  &  $2.30^{0.97}_{0.66}$  \\
\noalign{\smallskip}

  \hline
  \end{tabular}
  \begin{tablenotes}\footnotesize
  \item \emph{Note} -- $^{(\mathrm{a})}$ Fixed to zero. $^{(\mathrm{b})}$ $q_1$ and $q_2$ as defined by \citet{kipping13}. $^{(c)}$ Calculated from the scaled parameters as described by \citet{winn10}. $^{(d)}$~Assuming albedo = 0. 
\end{tablenotes}
\end{center}
\end{table*}

\end{acknowledgements}

\bibliography{gj9827} 
\bibliographystyle{apalike}

\begin{appendix}

\begin{table*}
\begin{center}
\caption{FIES RV measurements of GJ\,9827}
\begin{tabular}{lcccccccc}
\hline
\noalign{\smallskip}
$\rm BJD_{TDB}^1$ & RV  & $\pm \sigma$ &    BIS  & FWHM   & Ca\,II S index & $\pm \sigma$ & T$_\mathrm{exp}$ & S/N $^2$\\
-2450000          & (\kms)  & (\kms)       &  (\kms) & (\kms) &                &              &      (s)         &   \\
\hline
\noalign{\smallskip}
\multicolumn{4}{l}{FIES} \\
\noalign{\smallskip}
7954.617085 & 31.7746 & 0.0033 &      -  &     -  &    -   &    -  &   2700 & 55.2  \\
7955.612895 & 31.7724 & 0.0032 &      -  &     -  &    -   &    -  &   2700 & 56.0  \\
7956.627456 & 31.7751 & 0.0025 &      -  &     -  &    -   &    -  &   2700 & 68.5  \\
\dotfill & \dotfill & \dotfill &      -  &     -  &    -   &    -  &   \dotfill & \dotfill  \\
\noalign{\smallskip}
\hline
\end{tabular}
\label{tabRV1}
\\
\end{center}
\end{table*}

\begin{table*}
\begin{center}
\caption{HARPS RV measurements of GJ\,9827}
\begin{tabular}{lcccccccc}
\hline
\noalign{\smallskip}
$\rm BJD_{TDB}^1$ & RV  & $\pm \sigma$ &    BIS  & FWHM   & Ca\,II S index & $\pm \sigma$ & T$_\mathrm{exp}$ & S/N $^2$\\
-2450000          & (\kms)  & (\kms)       &  (\kms) & (\kms) &                &              &      (s)         &   \\
\hline
\noalign{\smallskip}
\multicolumn{4}{l}{HARPS} \\
\noalign{\smallskip}
7984.653428 & 31.9468 & 0.0013 &  0.0603 & 6.1447 &  0.679 & 0.010 & 2000 &  79.9 \\
7984.773491 & 31.9481 & 0.0016 &  0.0611 & 6.1409 &  0.662 & 0.016 & 1800 &  64.8 \\
7984.843042 & 31.9467 & 0.0018 &  0.0586 & 6.1466 &  0.630 & 0.021 & 3600 &  60.3 \\
\dotfill & \dotfill & \dotfill &  \dotfill & \dotfill &  \dotfill & \dotfill & \dotfill &  \dotfill \\
\noalign{\smallskip}
\hline
\end{tabular}
\label{tabRV2}
\\
\end{center}
\end{table*}

\begin{table*}
\begin{center}
\caption{HARPS-N RV measurements of GJ\,9827}
\begin{tabular}{lcccccccc}
\hline
\noalign{\smallskip}
$\rm BJD_{TDB}^1$ & RV  & $\pm \sigma$ &    BIS  & FWHM   & Ca\,II S index & $\pm \sigma$ & T$_\mathrm{exp}$ & S/N $^2$\\
-2450000          & (\kms)  & (\kms)       &  (\kms) & (\kms) &                &              &      (s)         &   \\
\hline
\noalign{\smallskip}
\multicolumn{4}{l}{HARPS-N} \\
\noalign{\smallskip}
7963.592670 & 31.9498 & 0.0014 &  0.0478 & 6.1011 &  0.700 & 0.010 & 1800 &  69.3 \\
7965.613121 & 31.9496 & 0.0012 &  0.0456 & 6.1111 &  0.714 & 0.008 & 1800 &  77.9 \\
7965.691320 & 31.9531 & 0.0012 &  0.0447 & 6.1120 &  0.746 & 0.008 & 1800 &  75.8 \\
\dotfill & \dotfill & \dotfill &  \dotfill & \dotfill &  \dotfill & \dotfill & \dotfill &  \dotfill \\
\noalign{\smallskip}
\hline
\end{tabular}
\label{tabRV3}
\\
\end{center}
Notes:\\
$^1$ Barycentric Julian dates are given in barycentric dynamical time. \\
$^2$ S/N per pixel at 550 nm. \\
\end{table*}

\end{appendix}

\end{document}